\documentclass[a4paper,11pt]{article}
\usepackage{jheppub} 
\usepackage{lineno}
\usepackage{enumerate}

\usepackage{definitions}%

\usepackage[T1]{fontenc} 
\usepackage{braket}
\usepackage[normalem]{ulem}
\usepackage{graphicx}
\usepackage{mathrsfs}
\usepackage{amsmath, amssymb}
\usepackage{bbold}
\usepackage{comment}
\excludecomment{confidential}
\usepackage{makecell}
\usepackage{bm}
\usepackage{braket}
\usepackage{slashed}
\usepackage{mathtools}

\usepackage{amsthm}

\newcommand{\mz}[1]{{#1}}
\newcommand{\mzo}[1]{{#1}}
\newcommand{\mzb}[1]{{#1}}
\newcommand{\mzc}[1]{{#1}}

\newcommand{\Ll}[1]{\overleftarrow{#1}}
\newcommand{\Rr}[1]{\overrightarrow{#1}}

\title{Generalized Wigner-Weyl calculus for gauge theory and non-dissipative transport}

\author{Praveen D. Xavier,}

\emailAdd{praveen.xavier@msmail.ariel.ac.il}

\author{M.A. Zubkov}
\affiliation{Physics Department, Ariel University, Ariel 40700, Israel}
\emailAdd{mikhailzu@ariel.ac.il}

\abstract{
We consider a theory of fermions interacting with a (in general, non-Abelian) gauge field. The theory is assumed to be essentially inhomogeneous, which might be provided by non-trivial  background fields interacting with both fermions and gauge bosons. For this theory, a version of Wigner-Weyl calculus is \mzc{developed}, in which the Wigner transformation of the fermion Green function \mzo{belongs to a matrix representation of gauge group}. We demonstrate the power of the proposed formalism through the representation of  responses of vector and axial currents to the gauge field strength through the topological invariants composed of the Wigner transformed two-point Green functions. This way a new family of non-dissipative transport phenomena is introduced. In particular, we discuss the non-Abelian versions of the chiral separation effect and of the quantum Hall effect.
}

\makeatletter
\gdef\@fpheader{}
\makeatother

\begin{document}
\maketitle
\flushbottom

\section{Introduction}

There is a long history of using topological invariants in quantum field theory with their applications to both high energy physics and condensed matter physics. Possibly, the most well-known example is the description of the Quantum Hall Effect (QHE) via the TKNN invariant \cite{Thouless1982}. This invariant has been proposed for homogeneous systems in the presence of constant magnetic field. It is also relevant for the intrinsic anomalous QHE in homogeneous topological insulators without external magnetic field. The value of the TKNN invariant is robust to smooth changes of the system   \cite{Avron1983,Fradkin1991,Hatsugai1997,Qi2008,Kaufmann:2015lga,Tong:2016kpv}. However, the TKNN invariant, by construction, is given by the integral of the Berry curvature over the occupied energy levels and, as such, it is neither defined for systems with interactions, nor for the case when an inhomogeneity is present. This gap has been filled by more involved constructions which include topological invariants composed of Green functions. The first step towards this direction has been made for the description of the intrinsic AQHE in $2D$ topological insulators  \cite{IshikawaMatsuyama1986,Volovik1988} (see also Chapter 21.2.1 in \cite{Volovik2003a}) where the Hall conductivity is given by (in units of $e^2/\hbar$)
\begin{equation}
\sigma_H = \frac{\cal N}{2\pi} \label{QHEI}
\end{equation}
with
{\be
	{\cal N}
	= \mz{-} \frac{ \epsilon_{ijk}}{  \,3!\,4\pi^2}\, \int d^3p \Tr
	\[
	{G}(p ) \frac{\partial {G}^{-1}(p )}{\partial p_i}  \frac{\partial  {G}(p )}{\partial p_j}  \frac{\partial  {G}^{-1}(p )}{\partial p_k}
	\].
	\label{N-0}
	\ee}%
Here $G(p)$ is the two-point Green function in momentum space. 
Topological invariants similar to that of  (\ref{N-0}) were considered also in condensed matter physics in the description of $3D$ fermion systems with a Fermi surface. Namely, the  topological invariant  responsible for the stability of the Fermi surface has the form:
$
N_1= \tr \oint_C \frac{1}{2\pi \ii}
G(p_0,p)d G^{-1}(p_0,p)$, where  $C$ is the closed path in Euclidean momentum space around the Fermi surface \cite{Volovik2003a} in momentum space. The stability of Fermi points is protected by a topological invariant, which has a form even closer to that of (\ref{N-0}) 
 \cite{Matsuyama1987a,Volovik2003a}:
\be
N_3=\frac1{24\pi^2} \epsilon_{\mu\nu\rho\lambda} \tr\int_S dS^\mu
G\partial^\nu G^{-1}
G\partial^\rho G^{-1}
G\partial^\lambda G^{-1}.
\label{N3}
\ee
In this expression $S$ is the three-dimensional hypersurface in $4D$ momentum space. It encloses the  Fermi point, and may be infinitely small. Similar topological invariants were used in other chapters of  condensed matter physics theory \cite{HasanKane2010,Xiao-LiangQi2011,Volovik2011,Volovik2007,VolovikSemimetal}. In particular, gapless fermions existing at the edges of topological insulators are protected by such invariants due to the corresponding bulk-boundary correspondence \cite{Gurarie2011a,EssinGurarie2011}. Gapless fermions in the bulk of topological semimetals \cite{Volovik2003a,VolovikSemimetal} are protected by similar invariants. The same is true of fermion zero modes living in the core of topological defects in the $^3$He-superfluid \cite{Volovik2016}. In high energy physics theory, similar topological invariants in momentum space have been discussed as well \cite{NielsenNinomiya1981a,NielsenNinomiya1981b,So1985a,IshikawaMatsuyama1986,Kaplan1992a,Golterman1993,Volovik2003a,Hovrava2005,Creutz2008a,Kaplan2011}.

Although the TKNN invariant and its form in \eqref{N-0} were originally proposed for systems without interactions and without inhomogeneity, it was widely believed that both these expressions remain relevant for models with weak interactions and even in the presence of disorder. The former statement for  (\ref{N-0}) has been proven within $2+1$D Quantum Electrodynamics \cite{ColemanHill1985,Lee1986}. This proof has been extended to a more wide class of {\it homogeneous} models in \cite{ZZ2019_0}.

For QHE systems in the presence of constant homogeneous magnetic field the role of interactions in the description of Hall conductivity has been discussed in many publications \cite{KuboHasegawa1959,Niu1985a,Altshuler0,Altshuler}. It has been found that the QHE conductivity remains robust to the introduction of weak interactions. The case when both interactions and disorder are present is more involved. 
It was considered in \cite{ZZ2019_1,ZZ2021,ZW2019,SZZ2020}, where it was found that the Hall conductivity averaged over the system area is given by (\ref{QHEI}) with 
\be
{\cal N}
=  \mz{-} \frac{T \epsilon_{ijk}}{ |{\bf A}| \,3!\,4\pi^2}\, \int \D{{}^3x} \int_{\cM}  \D{{}^3p}
\, {\rm tr}\, {G}_{W}(x,p )\star \frac{\partial {Q}_{W}(x,p )}{\partial p_i} \star \frac{\partial  {G}_{W}(x,p )}{\partial p_j} \star \frac{\partial  {Q}_{W}(x,p )}{\partial p_k}
\label{calM2d230I}
\ee
In this expression $T$ is temperature (assumed to be small), $|{\bf A}| $ is the system area, ${G}_{W}(x,p )$ is the Wigner transform of the one-particle electron Green function  $\hat G$, and the operator $\hat{Q}$ is the inverse of $\hat G$. Correspondingly, $Q_W $ is the Weyl symbol of $\hat Q$. By the star we denote the Moyal product. 
 (\ref{calM2d230I}) has been obtained for lattice models using the so-called approximate lattice Wigner-Weyl calculus, which works for the case of weak inhomogeneity: when external fields vary slowly at the distance of the order of the lattice spacing. 

Wigner-Weyl calculus itself, in its original form, stemmed from the works of H. Weyl \cite{Weyl1927} and E. Wigner \cite{Wigner1932}. It was developed in works by H. Groenewold \cite{Groenewold1946} and J. Moyal \cite{Moyal1949} in the context of ordinary quantum mechanics. Originally the main idea was to use the Wigner distribution instead of the wave function, and replace operators of physical observables by their Weyl symbols (defined as functions in phase space). The operator products are then replaced by the Moyal (star) products of Weyl symbols \cite{Ali2005,Berezin1972}. Many applications of this formalism to various problems in quantum mechanics were proposed  \cite{Curtright2012,Zachos2005} -- the most well-known of those is the case of anharmonic oscillator. Being modified accordingly, Wigner-Weyl calculus has been applied subsequently to high energy physics theory as well as to condensed matter physics \cite{Cohen1966,Agarwal1970,E.C.1963,Glauber1963,Husimi1940,Cahill1969,Buot2009}.
In particular this calculus has been used in QCD \cite{Lorce2011,Elze1986}, in quantum kinetic theory \cite{Hebenstreit2010,Calzetta1988}, and within non-commutative field theories \cite{Bastos2008,Dayi2002}. Several applications have been found to a variety of quantum systems and even to cosmology \cite{Habib1990,Chapman1994,Berry1977}.

While a majority of original works were devoted to systems defined in continuous spacetime, applications to solid state physics require the use of lattice models. Development of Wigner-Weyl calculus for lattice models encountered several problems. Attempts to overcome these problems began a long time ago \cite{Schwinger570}. The important results in this field of research resulted from the works of F.Buot \cite{Buot1974,Buot2009,Buot2013}, Wooters \cite{WOOTTERS19871}, Leonhardt \cite{Leonhardt1995}, Kasperowitz \cite{KASPERKOVITZ199421}, Ligab\'o \cite{Ligabo2016}. It is worth mentioning that Wigner-Weyl calculus is intimately related to deformation quantization  \cite{BAYEN197861,Kontsevich2003,Felder2000,Kupriyanov2008}. 

The approximate version of lattice Wigner-Weyl calculus  \cite{ZZ2019_1,ZZ2021,ZW2019,SZZ2020} (see also references therein) has been successful in applications to several practical problems related to non-dissipative transport phenomena \cite{Zubkov2017,Chernodub2017,Khaidukov2017,Zubkov2018a,Zubkov2016a}. The general output of these studies is that the response of non-dissipative currents to the corresponding external fields is given by topological invariants which are robust to smooth modifications of the systems. 

A precise version of the Wigner-Weyl formalism for lattice models suited for the description of the QHE has been developed recently \cite{FZ2019,Z2022}. Within this formalism it is possible to consider the lattice models precisely -- without approximation of slow varying external fields. In particular, this formalism allows one to deal with artificial lattices, in which magnetic flux through the lattice cell approaches the order of magnitude of the magnetic flux quantum, and, correspondingly, the Hofstadter butterfly might be observed.   

Investigation of non-dissipative transport effects has a long history. These effects exist within condensed matter physics and high energy physics \cite{Metl,Kharzeev:2013ffa,Kharzeev:2015znc,Kharzeev:2009mf,ref:HIC,9,Landsteiner:2012kd,semimetal_effects6,Gorbar:2015wya,Miransky:2015ava,Valgushev:2015pjn,Buividovich:2015ara,Buividovich:2014dha,Buividovich:2013hza}. 
One of such effects is the chiral separation effect (CSE) proposed in  \cite{Metl}. Its essence is the appearance of axial current directed along the external magnetic field in the presence of non-zero chemical potential in a system of relativistic Dirac fermions. For massless fermions the axial current in these systems is proportional to the external magnetic field strength $F_{ij}$ and to the ordinary chemical potential $\mu$. 
It is expected that this effect exists in the quark-gluon plasma (QGP) phase of quark matter existing at large temperatures and may be observed experimentally in heavy ion collisions  \cite{Kharzeev:2015znc,Kharzeev:2009mf,Kharzeev:2013ffa,ref:HIC} where the quark-gluon plasma appears in the presence of  large magnetic fields \cite{QCDphases,1,2,3,4,5,6,7,8,9,10}.

As it was mentioned above, the response of non-dissipative currents to external fields (i.e. their conductivities) are in many cases expressed through topological invariants. Quark matter  at nonzero baryon chemical potential is an interesting example of such a system, in which -- for technical reasons -- the lattice numerical simulation cannot be used, while   perturbative calculations also cannot give the complete picture of physical phenomena \cite{QCDphases,1,2,3,4,5,6,7,8,9,10}. Here we are in a situation when the existing computational methods fail to describe dynamics properly. In this situation the non-dissipative transport phenomena (in particular, the chiral separation effect) can still be investigated analytically. Namely, their conductivities, being topological invariants, are robust to smooth modification of the system. Therefore, if a simple model is found that is connected to real QCD by smooth modification (without phase transition on the way), the conductivity of interest may be calculated within such a model. Namely, in \cite{SZ2020,ZA2023} it was shown that in the presence of chiral symmetry the response of axial current (averaged over the system volume $\bf V$) to the external magnetic field and to the chemical potential is given by 
\begin{equation}
	\frac{d}{d\mu}\bar{J}_5^k = \frac{\mathcal N}{4\pi^2}\epsilon^{\mz{0ijk}}  F_{ij}\label{1_}
\end{equation}
where $\mathcal{N}$ is a topological invariant expressed through the Wigner transformed Green function $G_W$ and Weyl symbol of the Dirac operator $Q_W$:
\begin{eqnarray}
	\mathcal{N}
	&=&\frac{1}{48 \pi^2 {\bf V}}
	\int d^3x \int_{\Sigma_0}
	\tr \Bigg[\gamma^5
	{G}_W\star d { Q}_W \star { G}_W
	\wedge \star d { Q}_W\star { G}_W \star \wedge d { Q}_W
	\Bigg]\label{Ncompl0}
\end{eqnarray}
\mzb{The hypersurface $\Sigma_0$ in $4D$ momentum space consists of the two hyperplanes $p_4 = \pm \epsilon \to 0$.  The singularities of expression standing inside the integral are situated between these two hyperplanes. $\gamma^5$ commutes or anticommutes with $Q_W$ and $G_W$ in a vicinity of the positions of these singularities (the latter represent extension of the notion of Fermi surface to the non - homogeneous systems).  With such a definition
 (\ref{Ncompl0}) is a topological invariant. }

It is expected that in quark matter there is a phase transition at $T=0$ and finite $\mu$, but the properties of the hypothetical phases situated there are not known \cite{Simonov2007jb}. There may exist at large $\mu$ color superconductivity, the quarkyonic phase or a phase with restored chiral symmetry. In the latter case the results of \cite{ZA2023} may be applied directly to the investigation of the CSE in this phase. This part of the phase diagram may be relevant, in particular, for the description of physics inside neutron stars~\cite{Cook:1993qr}. 

There is an intimate relation between the chiral anomaly and the CSE \cite{Zyuzin:2012tv} as well as the chiral magnetic effect  (CME) \cite{Vilenkin,CME,Kharzeev:2013ffa,Kharzeev:2009pj,SonYamamoto2012}. While in thermal equilibrium the CME is absent \cite{Valgushev:2015pjn,Buividovich:2015ara,Buividovich:2014dha,Buividovich:2013hza,Z2016_1,nogo,nogo2,BLZ2021}, it is present out of equilibrium -- even very close to equilibrium \cite{BLZ2022}. Moreover, the CME also manifests itself in the magneto-conductivity of relativistic fermionic matter 
 \cite{Nielsen:1983rb} and may be observed experimentally in Weyl and Dirac semimetals \cite{ZrTe5}.   

An important arena for the experimental observation of non-dissipative transport phenomena is Dirac and Weyl semimetals \cite{semimetal_effects6,semimetal_effects10,semimetal_effects11,semimetal_effects12,semimetal_effects13,Zyuzin:2012tv,tewary,16}.
In these materials emergent relativistic invariance is present, and the Dirac fermions emerge. The CSE may be observed experimentally in magnetic Weyl semimetals through the observation of the so-called Weyl orbits \cite{Z2024}.

In the above mentioned cases of QHE and CSE the conductivities are expressed through topological invariants using Wigner-Weyl calculus. 
In both cases it is the response of certain non-dissipative currents to the {\it Abelian} field strength.  
The present paper is devoted to the extension of the two mentioned effects  (QHE and CSE) to the case when the external gauge field is {\it non-Abelian}. Correspondingly, we generalize the Wigner-Weyl calculus of \cite{ZZ2021,ZA2023,Z2024} (see also references therein) to a form which is suitable for dealing with non-Abelian gauge fields. \mzc{We accept the definition of covariant Weyl symbol of operator proposed long time ago in \cite{Vasak:1987um}. This Weyl symbol transforms covariantly under the gauge transformation unlike the one adopted in \cite{ZZ2021,ZA2023,Z2024}. This allows us to performed the further analysis of non - Abelian gauge theory (rather than the Abelian one) along the lines of \cite{ZZ2021,ZA2023,Z2024}. Notice that the Weyl symbol of  \cite{Vasak:1987um} (see also references therein) was used widely in certain other applications, in particular, during extension of the approach to the systems in curved space - time \cite{Antonsen_1997}, and to the chiral kinetic theory \cite{chen2013berry}.}

\section{Statement of the main results}

\subsection{Generalized \mzc{(or, covariant)} Wigner-Weyl calculus}

\label{SectMain1}

We are working with a quantum field theory defined in continuous $D$-dimensional Euclidean spacetime. 
Consider the quantum mechanical operator 
\begin{equation}
	\hat{ \pi}_\mu:=\hat p_\mu -  A_{\mu}(\hat x)
\end{equation}
where $A_{\mu}(\cdot)\equiv A_{\mu}^a(\cdot) {t}^a$ is an external non-Abelian gauge field and $\hat x^\mu, \hat p_\mu$ are the $D$-dimensional quantum mechanical position and momentum operators \mzc{with canonical commutation relations $[\hat x^\mu, \hat p_\nu] = i \delta^{\mu}_{\nu} $ (we are using the system of units with $\hbar = c = 1$). The operators to be considered below are defined in the Hilbert space $\cal H$ spanned on the eigenvectors of operators $\hat x^\mu$ or $\hat p_\mu$ denoted correspondingly by ket - vectors $\ket{x}$ or $\ket{p}$. Conjugated space ${\cal H}^*$ is spanned on bra - vectors $\bra{x}$ or $\bra{p}$. The scalar products are $\braket{x|x'} = \delta^{(D)}(x-x') $, $\braket{p|p'} = \delta^{(D)}(p-p') $, $\braket{x|p} = \braket{p|x}^* = \frac{1}{(2\pi)^{D/2}}e^{i p x} $. Thus we deal with the mathematical basis of ordinary quantum mechanics in $D$ - dimensional space. Mostly we will consider the case $D=4$ while the conventional quantum mechanics in $3+1$ D space - time deals with $D =3$.  }


 We define the generalized Weyl symbol (we will call it also the generalized Wigner transform or simply `the Wigner transform') of an operator $\hat{{X}}$ as
\begin{equation}
	X_{{W}}(x,p)=\int dy~e^{ipy}\bra{x}e^{-\frac{i}{2}y\hat{ {\pi}}}\hat{  X} ~e^{-\frac{i}{2}y\hat{ {\pi}}}\ket{x} \label{10}
\end{equation}
\mzc{This definition has been proposed for the first time in \cite{Vasak:1987um} and was used since then in many applications. We are going to propose the specific application of this Weyl symbols intended to derive the topological expressions of certain non - dissipative currents.}
In general, $\hat{X}$ and $X_{{W}}$ are matrices with the matrix dimensions of $A$.
 \eqref{10} can also be written as follows:
\begin{equation}
	X_{{W}}(x,p)=\int dy~e^{ipy}  U(x , x-y/2)\bra{x-y/2}\hat{X}\ket{x+y/2}  U(x+y/2 , x)  \label{12}
\end{equation}
\mzc{Here \begin{equation}
		U(w,x)=\text{Pexp}\left(i\int_{x\to w} dz ~  A(z)\right)
	\end{equation}
	is the path-ordered exponential along the straight path from $x$ to $w$.}
From this expression it is obvious that $X_{{W}}(x,p)$ is transformed under the gauge transformation $\Omega(x)$ as $X_{{W}}(x,p)\to \Omega(x) X_{{W}}(x,p) \Omega(x)^\dagger$, i.e. that it belongs to the adjoint representation.
Besides, it is easy to prove the following expression for the trace of an operator:
\begin{equation}
	{\rm Tr}\, \hat{X} = \frac{1}{(2\pi)^D} \mzc{\text{tr}_D \, \text{tr}_G}\,\int dp dx X_W(x,p)\label{TrX}
\end{equation}
\mzc{Here and below by $\text{tr}_D$, $\text{tr}_G$ and $\text{tr}_H$ we denote the traces w.r.t. the spinor indices (and the other internal indices, excluding the gauge ones), gauge indices and the Hilbert space respectively. Obviously 
$$ 
{\rm Tr} = \text{tr}_D \, \text{tr}_G \, \text{tr}_H,
$$
 and Eq. (\ref{TrX}) replaces the trace with respect to  Hilbert space by an integral in phase space.}

By the star product of the Weyl symbols of two operators we understand the Weyl symbol of the products of these operators:
\begin{align}
	\begin{split}
		&(\hat{  X}\hat{ {Y}})_{{W}}(x,p)=X_{{W}}(x,p)\bigstar Y_{{W}}(x,p) \label{yoni00}
	\end{split}
\end{align}

\mzc{Below we summarize the main properties of the generalized Weyl symbol (or, covariant Weyl symbol) introduced above, and of the corresponding star product.}

\begin{enumerate}

\item

	The inverse (Weyl) transform is
	\begin{equation}
		\hat{ {X}}=(2\pi)^{-2D}\int dq dy dx dp ~e^{\frac{i}{2}y(\hat{ {\pi}}-p)}e^{iq(\hat x-x)} {X}_{{W}}(x,p)~e^{\frac{i}{2}y(\hat{ {\pi}}-p)} \label{2}
	\end{equation}

\item
 
	\begin{align}
		\begin{split}
			&X_{{W}}\bigstar Y_{{W}}\\
			&=(2\pi)^{-4D}\int dw ~e^{ipw}\int dqdydzdk dq'dy'dz'dk'~e^{-i(yk+qz+y'k'+q'z')}\\
			&\quad\times \bra{x}e^{-\frac{i}{2}w\hat{ \pi}}e^{\frac i 2 y\hat{  {\pi}}}e^{iq\hat x}  X_{{W}}(z,k)e^{\frac i 2 y\hat{  {\pi}}}e^{\frac i 2 y'\hat{  {\pi}}}e^{iq'\hat x}  Y_{{W}}(z',k')e^{\frac i 2 y'\hat{  {\pi}}}e^{-\frac{i}{2}w\hat{ \pi}}\ket{x}\\
			&=(2\pi)^{-2D}\int dydk dy'dk'~e^{-iy(k-p)-iy'(k'-p)}\\
			&\quad\times  {  U(x , x-(y+y')/2)  U(x-(y+y')/2 , x-y'/2)  X_{{W}}(x-y'/2,k)   U(x-y'/2 , x+(y-y')/2)}\\
			&\quad\quad {  U(x+(y-y')/2, x+y/2)  Y_{{W}}(x+y/2,k')  U(x+y/2,  x+(y+y')/2)  U(x+(y+y')/2 , x)}  \label{yoni0}
		\end{split}
	\end{align}

\item

	\begin{align}
		\begin{split}
			&{  X}_W(x,p) \bigstar { {Y}}_W(x,p) =\\&\bigg(e^{\frac{i}{2}(\Rr\partial_{p_1}+\Rr\partial_{p_2})\Rr D_{x}}e^{-\frac{i}{2}\Rr\partial_{ {p_1}}\Rr D_{x}} {X}_W(x,p_1)e^{-\frac{i}{2}\Rr D_{x} {\Ll\partial_{p_1}}}\\\
			&e^{-\frac{i}{2} {\Rr\partial_{p_2}}\Rr{D}_{x}} {Y}_W(x,p_2)e^{-\frac{i}{2}\Ll\partial_{ {p_2}}\Rr{D}_{x}}e^{\frac{i}{2}(\Ll\partial_{p_1}+\Ll\partial_{p_2})\Rr{D}_{x}}\bigg)\times 1\,\bigg|_{\substack{p_1=p_2=p}} \label{yo2}
		\end{split}
	\end{align}

\item

	For any operators $\hat{X}$ and $\hat{Y}$ we have 
	\begin{equation}
	\text{tr}_G	\text{tr}_H\, \hat{X} \hat{Y} = \text{tr}_G\frac{1}{(2\pi)^D} \int dp dx X_W(x,p)\bigstar Y_W(x,p) = \text{tr}_G\frac{1}{(2\pi)^D} \int dp dx X_W(x,p) Y_W(x,p) 
	\end{equation}
provided that the above integrals are convergent. Here $\text{tr}_G$ and $\text{tr}_H$ are traces w.r.t. the  gauge indices and the Hilbert space respectively.

\item

Let  $D_\mu\equiv\partial_\mu-iA_\mu$ be the gauge covariant derivative and consider the operator $Q(x,-iD)$ of the form of 
\mzo{
\begin{equation}
Q(x,-iD) = \sum_{|\alpha| \le m} o_{\alpha}(x) \circ (-i D)^\alpha \label{sym0}
\end{equation}
with integer number $m$ and the multi-index $\alpha=(\alpha_1,\alpha_2,\alpha_3,\alpha_4)$ and $|\alpha| = \sum_i \alpha_i$. By $(-i D)^\alpha$ we denote the product of derivatives 
$(-i D)^\alpha = \Pi_\mu (-i D)_\mu^{\alpha_\mu}$. By $o_{\alpha}(x) \circ (-i D)^\alpha$ we denote 
\begin{equation}
	o_{\alpha}(x) \circ (-i D)^\alpha = \frac{1}{2^{|\alpha|}} \{...\{o_{\alpha}(x),(-i D_1)\}...(-i D_1)\}(-i D_2)\}...(-i D_2)\} ...(-i D_4)\}
\end{equation}
Here $ (-i D_i)$ is encountered $\alpha_i$ times. For each $\alpha$  $o_{\alpha}(x)$ is a matrix - valued function of coordinates. This is matrix both in spinor space and in internal space, but it is singlet in gauge group of the gauge field $A$. 
 We assume that the given operator is Hermitian, and $\{Q(x,-iD),\gamma^5\}=0$. In chiral representation of gamma matrices we have 
 \begin{equation}
 Q(x,-iD) = \left(\begin{array}{cc}
0 & \hat{O} \\ \hat{O}^\dagger & 0
 \end{array}\right)
 \end{equation}
 We require that $\hat{O}$ is Elliptic operator. }

 We also denote  $\hat Q:=Q(\hat x,\hat \pi)$. Then the following exact relationship holds:
	\begin{equation}
		Q_{{W}}(x,p)\equiv(\hat Q)_{{W}}(x,p)=Q(x,p) \label{exact}
	\end{equation} 
\mzo{Here $Q(x,p)$ is understood as
	\begin{eqnarray}
	Q(x,p)&=&\sum_{|\alpha| \le m} o_{\alpha}(x)  p^\alpha
\end{eqnarray}
that is we simply substitute to Eq. (\ref{sym0}) $p$ instead of $-i D$ in order to obtain the above expression \footnote{\mzo{We can also consider the more complicated situation, when operator  is defined through the formal series
	$$
		Q(x,-iD)={\rm lim}_{m \to \infty}\sum_{|\alpha| \le m} o_{\alpha}(x) \circ (-i D)^\alpha
	$$
	In order to give the precise meaning to the above formal series let us reorder using conventional commutation relations of these operators the terms in the $\circ$  - product in such a way that all functions of $x^{\nu_i}$ (including $A(x)$, and its derivatives) stand on the left while ordinary derivatives $\partial_{\mu_j}$ stand on the right: 
	$$
		Q(x,-iD)=\sum_{\alpha}  \tilde{o}_{\alpha}(x)(-i\partial)^\alpha \label{sym1}
	$$
	Here $\tilde{o}_{\alpha}(x))$ is tensor that is made of ${o}_{\alpha}(x)$, and $A(x)$ with its derivatives using  commutation relations between  $A(x)$, and  $D_{\mu_j}$. Now we give the meaning to these formal series as follows. We determine the mixed matrix elements
	$$
		\bra{x}Q(x,-iD)\ket{p}=\sum_{\alpha}  \tilde{o}_{\alpha}(x)p^\alpha
	$$
	This series is understood as the definition of analytical function of $x^\nu$ and $p_\mu$ in a certain (possibly, very small) vicinity of  $p=0$ that is we assume that the given series is convergent in a certain vicinity of this point in momentum space. The whole function at the arbitrary values of $x$ and $p$ is then understood as its analytical continuation.} }.  
}


\end{enumerate}

\subsection{Wigner-Weyl gauge theory}
\label{SectMain2}
In 4 - dimensional Euclidean spacetime, consider an external, time-independent non-Abelian gauge field, $A_\mu(x)$, (with gauge group $G$), minimally coupled to fermions (in the fundamental representation), where the partition function is
\begin{align}
	Z&=\int D[\bar\psi,\psi]~e^S \label{pf0} \\
	S&=-\int d^4x ~\bar\psi(x)Q(x,-iD)\psi(x)\nonumber
\end{align}
where $D_\mu\equiv\partial_\mu-iA_\mu$ is the gauge covariant derivative and $Q(x,-iD)$ is defined by  (\ref{sym0}). 

\mzc{We will show that the Wigner transformed Green functions obey the following properties.}

\begin{enumerate}

	\item

	The Wigner transform of the Green function $G_W$ obeys the generalized Groenewold equation
	\begin{equation}
		G_W(x,p) \bigstar Q(x,p) = 1
	\end{equation}


\item

The non-Abelian vector current (which belongs to the adjoint representation of the gauge group) is defined as the Noether current corresponding to the gauge transformation 
\begin{align}
	\psi(x)\to e^{i\alpha(x)}\psi(x) \label{this} \\
	\bar\psi(x)\to\bar\psi(x)e^{-i\alpha(x)} \label{this1} 
\end{align}
where $\alpha(x)$ belongs to the Lie algebra of the gauge group. Recall that we denote by $\text{tr}_{D}$ the trace with respect to the indices other than those corresponding to the given gauge group.

In terms of the Wigner transform of the Green function, $G_W$, the average vector current is given by 
\begin{equation}
	\langle J_{\mu}(x)\rangle=-\text{tr}_{D}\int\frac{d^4p}{(2\pi)^4}G_W\partial_{p_\mu} Q \label{unrahv}
\end{equation}


\item

The axial current is defined as the Noether current corresponding to the axial gauge transformation
\begin{align}
	\psi(x)\to e^{i\alpha(x)\gamma^5}\psi(x) \label{this} \\
	\bar\psi(x)\to\bar\psi(x)e^{i\alpha(x)\gamma^5} \label{this1} 
\end{align}

The averaged axial current is given by 
\begin{equation}
	\langle J_{\mu}(x)\rangle=-\frac{1}{2}\text{tr}_{D}\int\frac{d^4p}{(2\pi)^4}G_W\partial_{p_\mu}[Q,\gamma^5]  \label{unrah0}
\end{equation}

\item

	To distinguish between the $x$ dependence arising from the inhomogeneity (caused by the background fields) vs. from the gauge field we introduce the function $G_{{W}}(x,p,z)$ (the $z$ dependence arises from the dependence of $\bigstar$ on the gauge field while the $x$ dependence is the consequence of the background inhomogeneity). 
	\mzc{Namely, $G_{{W}}(x,p,z)$  is the solution of 
		\begin{align}
			\begin{split}
				1=&(2\pi)^{-2D}\int dydk dy'dk'~e^{-iy(k-p)-iy'(k'-p)}Q(x-y'/2,k)\\
				&  U(z , z-(y+y')/2)  U(z-(y+y')/2 , z-y'/2)     U(z-y'/2 , z+(y-y')/2)\\
				&U(z+(y-y')/2, z+y/2)  G_{{W}}(x+y/2,k',z+y/2)  U(z+y/2,  z+(y+y')/2)  U(z+(y+y')/2 , z) \label{id10''}
			\end{split}
		\end{align}
		\eqref{id10''} is to be compared with \eqref{yoni0}. One can see that the difference is that in \eqref{id10''} the dependence of the parallel transporters on $x$ is replaced by the dependence on $z$. Therefore, it is obvious that $G_{{W}}(x,p)=G_{{W}}(x,p,x)$.}
	
	{Note that $Q(x,p)$ has no $z$ dependence.}
	The solution of the Groenewold equation may be expanded in powers of the covariant derivative $D_z$: 
	\begin{equation}
		G_{{W}}(x,p,z)=\sum_{n\geq 0} G^{(n)}(x,p,z).
	\end{equation}
	Here $G^{(n)}(x,p,z)$ contains $n$ powers of $D_z$. To lowest order in $D_z$ the Groenewold equation becomes
	\begin{equation}
		Q\star G^{(0)}=1 \label{bumb}
	\end{equation}
	where $\star = e^{\frac{i}{2}(\overleftarrow{\partial}_x\overrightarrow{\partial}_p-\overleftarrow{\partial}_p\overrightarrow{\partial}_x)}$ is the ordinary Moyal product.
	In the next orders we have $G^{(1)}=0$, and 
	\begin{align}
		G^{(2)}(x,p,z)&=-\frac{i}{2}G^{(0)}(x,p)\star\partial_{p_{\mu}}Q(x,p)\star \partial_{p_{\nu}}G^{(0)}(x,p)F_{\mu\nu}(z),\label{G2}
	\end{align}
where $F_{\mu\nu}$ is non-Abelian gauge field strength.


\end{enumerate}

\subsection{Non-Abelian non-dissipative transport}
\label{SectMain3}
We use Pauli-Villars regularization in order to \mzo{subtract from the observable quantities the unphysical contributions of ultraviolet region in momentum space. For the details of this regularization see, for example, \cite{slavnov1977pauli}.}  It substitutes the expression for the thermodynamic potential ($T$ is temperature) \mzo{${\Omega} =-T{\rm log}\, Z$ by 
\begin{equation}
	{\Omega}_{reg} = -T\sum_{i} C_i \,{\rm log} \, Z(M_i),	\label{logZM}
\end{equation}}
where $Z(M_i)$ is calculated as  (\ref{pf0}) with $Q$ substituted by 
$Q_{(M_i)}(i\partial_p, p - A(i\partial_p))$ such that the nonzero parameter $M_i$ provides the appearance of a gap (i.e. a mass for the fermion). Coefficients $C_i$ obey the conditions:
\begin{equation}
	\sum_i C_i = 0, \quad \sum_i C_i M_i^2 = 0,\quad C_0=1\label{cond}
\end{equation}
\mzo{These conditions provide, in particular, vanishing contributions of ultraviolet momentum space region  to the thermodynamical potential of Eq. (\ref{logZM}). This occurs because at $|p| \gg M$ the contributions of the Pauli Villars regulators are equal to the contribution of the original fermions. Their sum vanishes because $\sum_i C_i = 0$. At the same time the contributions of Pauli - Villars regulators from region with $|p| \ll M$ vanish, so that only those of the original fermions remain. Thus the procedure of Pauli - Villars regularization (in the absence of inter - fermion interactions) results in the subtraction of unphysical artifacts originated in ultraviolet within the non - regularized fermionic field theory. Notice that the same job is done, for example, by lattice regularization.    }

Let us define 
\begin{align}
	\begin{split}
		N_3(x)&=\mz{-}\frac{1}{48\pi^2}\int_{\Sigma_0(x)} \text{tr}_{D}\left({G}^{(0)}\star d{Q}\star \wedge d{G}^{(0)}\wedge d{Q} \right)
	\end{split}\label{N3x}
\end{align}
where \mzb{the hypersurface $\Sigma_0$ in $4D$ momentum space consists of the two hyperplanes $p_4 = \pm \epsilon \to 0$.  The singularities of expression standing inside the integral are situated between these two hyperplanes. The positions of these singularities generalize the notion of the Fermi surface to the non - homogeneous systems.
 By $dG^{(0)}$ and $dQ$ we denote the one - forms being the external derivatives of  $Q((\vec{x},0)(\vec{p},\pm \epsilon))$ and $G((\vec{x},0)(\vec{p},\pm \epsilon))$ correspondingly considred as functions of spatial momenta $\vec{p} = (p_1,p_2,p_3)$  with fixed value of $\vec{x}$.  }

\begin{enumerate}
	\item{The non-Abelian version of the CSE}
	
	Suppose that the $3+1D$ fermionic system with Fermi surface is in the presence of an external non-Abelian magnetic field (i.e. when only the components $F_{ij}(x)$ are nonzero with $i,j = 1,2,3$). We suppose that in a small vicinity of the Fermi surface, chiral symmetry is present. Then, provided that the field strength varies slowly compared to $N_3(x)$ we obtain, for the response of axial current to chemical potential:
	\begin{align}
		\begin{split}
			\frac{d}{d\mu}\bar{J}^{(5)}_i&= \frac{1}{4\pi^2}\epsilon_{ijk}N_3 F_{jk} \label{CSENAG0}
		\end{split} 
	\end{align}	
	Here $N_3$ is given by \begin{align}
		\begin{split}
			N_3&=\mz{-}\frac{1}{48\pi^2 V}\int d^3 x\int_{\mzb{\Sigma_0}} \text{tr}_{D}\left(G^{(0)}\star dQ\star \wedge dG^{(0)}\wedge dQ \right),
		\end{split}\label{N30}
	\end{align}
where $V$ is the overall volume of the system. \mzc{Notice that Eqs. (\ref{CSENAG0}) and (\ref{N30}) are precisely the same for non - Abelian and Abelian cases. The only difference is the meaning of the field strength and of the axial current (Abelian or non - Abelian). In the Abelian case this representation has been obtained previousely (see \cite{ZA2023} and references therein). Now we generalize this result to the case of the non - Abelian gauge fields.  In Sections \ref{case1} and \ref{case2} we consider the particular cases of systems, in which the value of $N_3$ is equal to the number of Dirac fermions interacting with the given non - Abelian gauge field. In general case $N_3$ is a topological invariant, i.e. it is robust to smooth modifications of the system. This allows to calculate it for rather complicated systems, if it is known that they may be transformed smoothly to the simpler ones.  } 

\mzo{Notice that in the case of an inhomogeneous system the notion of conventional Fermi surface strictly speaking is not defined. The same refers also to the systems with interactions. For the case of an interacting system the notion of the Fermi surface can be defined, however, as the position of the poles of the Green function. The extension of the notion of the Fermi surface to the case of the system with inhomogeneity is also possible. In our present paper we consider the position (in momentum space) of the singularities of expression standing inside the integral in Eq. (2.32) as the position of the Fermi surface.  These positions form a two – dimensional surface in momentum space depending on the coordinate $x$. At each value of $x$ the three - dimensional hypersurface \mzb{$\Sigma_0$} embraces this two - dimensional "Fermi surface".  }

\mzo{As for the notion of the mass gap, it has the clear meaning in the systems we are considering. Namely, if for any value of chemical potential there exists the “Fermi surface” defined above, then there is no gap in the given system. If for the values of chemical potential within the interval between the two values $\mu_1$ and $\mu_2$ the “Fermi surface” is absent, then we can say that this interval represents the gap for the given system. }

\item{\mzc{The non - Abelian analogue of the quantum Hall effect (QHE) in the case of a \mzo{ $3+1$ D system}.} }

	Let us consider the case, when the fermion is massive, and in Euclidean spacetime its propagator does not have poles. Besides, we require that the external gauge field strength is varying slowly. Response of the spatial components of vector current (averaged over the system volume) to non-Abelian {\it electric} field strength (with nonzero $F_{4i} = - i E_i$ components) receives the form 
\begin{align}
	\begin{split}
		\bar{J}^{v,QHE}_i&=- \frac{1}{4\pi^2}\epsilon_{ijk}\tilde{M}^{*(v)j} E_{k} \label{QHEAG0}
	\end{split} 
\end{align}
with 
\mzc{\begin{align}
	\begin{split}
		\tilde{M}^{*(v)i}&= -\frac{1}{V \,24\pi^2}\Bigl[\int d^3 x \wedge dp^i\wedge  \text{tr}_{D}\left(G^{(0)}\star dQ\star \wedge dG^{(0)}\star\wedge dQ \right) \Bigr]_{reg}
	\end{split}\label{M30}
\end{align}
In this expression the integration is over the product of the spatial part of coordinate space ($d^3x \equiv dx^1 \wedge dx^2 \wedge dx^3$) and  the four - dimensional momentum space. The subscript "reg" means that the unphysical contribution of the Dirac sea should be subtracted from the considered quantity. Such a contribution does not present in \mzo{some} cases as demonstrated below, when the particular examples of the quantum systems are considered (Sect. \ref{SectEx}). However, if such a contribution is present the Pauli  - Villars regularization provides their cancellation.  } 
It is worth mentioning that  vector $\tilde{M}^{*(v)i}$ is related to the vector of intrinsic magnetic moment ${\cal M}$ of the fermion system (see Appendix \ref{Magnetic} ):
\mz{\begin{equation}
	\frac{\partial}{\partial \mu}{\cal M}_i = 	\frac{1}{4\pi^2} \tilde{M}^{*(v)i}
\end{equation} 
\mzc{As for the case of the CSE considered above, the expressions for the quantum Hall effect given by Eqs. (\ref{QHEAG0}) and (\ref{M30}) are identical for the Abelian and non - Abelian external gauge field. The Abelian case has been considered before (see \cite{Z2024CSE}). Now we generalize this result to the non - Abelian case using the developed formalism of covariant Wigner - Weyl calculus. Notice that strictly speaking Eq. (\ref{M30}) does not represent a topological invariant. However, in many cases it may be represented as an integral of a topological invariant (see below Sect. \ref{SectEx}). }
}

\item{\mzc{The non - Abelian analogue of the quantum Hall effect (QHE) in the case of a gapped $2+1$ D system.} }


Response of vector current (averaged over the system volume), in the gapped $2+1D$ system, to the non-Abelian electric field strength is (for the case of external field varying slowly)
\begin{align}
	\begin{split}
		\bar{J}^{v,QHE}_i&= \frac{1}{2\pi}\epsilon_{ij}M_3 E_{j} \label{QHEAG02}
	\end{split} 
\end{align}
with 
\mzc{\begin{align}
	\begin{split}
		M_3&=-\frac{1}{S \,24\pi^2}\Bigl[\int d^2 x\int \text{tr}_{D}\left(G^{(0)}\star dQ\star \wedge dG^{(0)}\star\wedge dQ \right) \Bigr]_{reg}
	\end{split}\label{M302}
\end{align}
Here $S$ is the area of the system, the integration is over the product of the two - dimensional spatial coordinate space and three - dimensional momentum space. As for the three - dimensional case in this expression the subscript "reg" means that the unphysical contribution of the Dirac sea should be subtracted. Eqs. (\ref{QHEAG02}) and (\ref{M302}) are precisely the same for ordinary (Abelian) QHE. For that case these results were obtained in \cite{ZW2020}. Now we generalize the result of \cite{ZW2020} to the case of the non - Abelian electric field and non - Abelian Hall current using the developed machinery of covariant Wigner - Weyl calculus. Notice that the above defined quantity $M_3$ is a topological invariant, which reflects the fact that the QHE conductivity (both Abelian and non - Abelian) is robust to the smooth modifications of the system.
}


\end{enumerate}

\mzc{Thus, summarized above are the main results of this paper.} Their proofs are given in the forthcoming sections.

\mzo{In the present paper we consider the response of axial and vector current to the external non – Abelian gauge field strength. At the same time, the system contains inhomogeneity of another source. We call the sources of this extra inhomogeneity “the background fields”. We do not consider response to these fields. In particular, our Dirac operator $Q$ entering Eqs. (\ref{N30}), (\ref{M30}), (\ref{M302}) (the expressions responsible for the CSE and QHE) contains these background fields, which provide its dependence on $x$. At the same time the external gauge field does not enter this $Q$. }

\section{Generalized Wigner-Weyl calculus for the gauge theory}
\label{Sect3}
\subsection{Generalized Wigner transform}
\mzc{We consider the theory in $D$ - dimensional Euclidean space with $A_{\mu}(\cdot)\equiv A_{\mu}^a(\cdot) {t}^a$ that is an external non-Abelian gauge field.  The construction used in the present paper is based on the notions of $\hat x^\mu, \hat p_\mu$ that are the $D$-dimensional position and momentum operators, and the covariant momentum operator $\hat{ \pi}_\mu=\hat p_\mu -  A_{\mu}(\hat x)$.  Definitions related to these operators and the corresponding Hilbert space of states are given above in Sect. \ref{SectMain1}. } 

\mzc{The generalized Weyl symbol or covariant Weyl symbol of an operator $\hat{{X}}$ is understood as the generalized/covariant Wigner transform of its matrix elements} (hereafter simply `the Wigner transform'):
\begin{equation}
      X_{{W}}(x,p)=\int dy~e^{ipy}\bra{x}e^{-\frac{i}{2}y\hat{ {\pi}}}\hat{  X} ~e^{-\frac{i}{2}y\hat{ {\pi}}}\ket{x} \label{1}
\end{equation}
In general, $\hat{X}$ and $X_{{W}}$ are matrices with the matrix dimensions of $A$. \mzc{Notice that the definition of Eq. (\ref{1}) has been proposed for the first time in \cite{vasak1987quantum}.}

Below we will prove that the inverse (Weyl) transform is
\begin{equation}
    \hat{ {X}}=(2\pi)^{-2D}\int dq dy dx dp ~e^{\frac{i}{2}y(\hat{ {\pi}}-p)}e^{iq(\hat x-x)} {X}_{{W}}(x,p)~e^{\frac{i}{2}y(\hat{ {\pi}}-p)} \label{2}
\end{equation}
\begin{proof}
Substituting \eqref{2} in the RHS of \eqref{1}, we obtain
\begin{equation}
    (2\pi)^{-2D}\int dy ~e^{ipy}\int dq'dy' dx'dp'\bra{x}e^{-\frac{i}{2}y\hat{ {\pi}}}e^{\frac{i}{2}y'\hat{ {\pi}}}e^{iq'\hat x} {X}_{{W}}(x',p')e^{\frac{i}{2}y'\hat{ {\pi}}}e^{-\frac{i}{2}y\hat{ {\pi}}}\ket{x}e^{-i(q'x'+y'p')} \label{1.4}
\end{equation}
Now we use the following fact\footnote{To prove this we note that 
\begin{equation}
    \exp(yD)=\lim_{N\to\infty}\prod_{n=1}^N (1+(y/N)D)
\end{equation}
But we can write 
\begin{align}
\begin{split}
    1+(y/N)D&=(1+(y/N)\partial_x)(1-i(y/N)A(x))\\
    &=\mathcal T({y/N})(1-i(y/N)A(x))
\end{split}
\end{align} (upto $\mathcal{O}(N^{-2})$ terms) where $\mathcal T({\varepsilon})$ is the translation operator: $\mathcal T({\varepsilon})\psi(x)=\psi(x+\varepsilon)$. So
\begin{equation}
    \exp(yD)\psi(x)=\lim_{N\to\infty}\left(\prod_{n=1}^N \mathcal T(y/N)(1-i(y/N)A(x))\right)\psi(x)
\end{equation}
which is easy to see gives the required result when we use the definition \eqref{yip} for $U(x,x+y)$.
} (see (3.10) of \cite{Vasak:1987um}):
\begin{equation}
    \exp(y D)\psi(x)
    =U(x,x+y)\psi(x+y), \quad \mz{\psi^{\dagger}(x)\exp(y \overleftarrow{D}^*)
    	=\psi^{\dagger}(x+y) U(x+y,x)} \label{yon}
\end{equation}
where $\psi(x)$ is any element of the {fundamental} representation and $D_{\mu}$ is the gauge covariant derivative: $D_\mu\equiv \partial_\mu-iA_\mu(x)$, {$D^*_\mu\equiv \partial_\mu+iA_\mu(x)$} and
\begin{equation}
        U(w,x)=\text{Pexp}\left(i\int_{x\to w} dz ~  A(z)\right)
\end{equation}
is the path-ordered exponential along the straight path from $x$ to $w$. Its definition is
\begin{equation}
      U(w,x)=\lim_{N\to\infty}  f_N(N)  f_N(N-1)...  f_N(0) \label{yip}
\end{equation}
where 
\begin{equation}
      f_N(n)=\left(1+\frac{i}{N}(w-x)^\mu  A_\mu\left(x+\frac{n}{N}(w-x)\right)\right)
\end{equation}
From the definition it is clear that $A(x)$ stands on the right, while $  A(w)$ stands on the left and that $  U$ is unitary and satisfies $  U(x,w)^{-1}=  U(w,x)$. 
If $\phi(x)$ lies in the adjoint representation instead then the analog of \eqref{yon} is
\begin{equation}
    \exp(y {\cal D})\phi(x) 
    =U(x,x+y)\phi(x+y)U(x+y,x) \label{yonez}
\end{equation}
{We denote here the covariant derivative in the adjoint representation by $\cal D$ to distinguish it from the one in the fundamental representation.}


In bra-ket notation \eqref{yon} reads
\begin{equation}
    e^{-iy\hat{ {\pi}}}\ket{x}=\ket{x+y} {U(x+y,x)}
\end{equation}
and 
 {\begin{equation}
\bra{x}	e^{iy\hat{{\pi}}}={  U(x , x+y)}\bra{x+y}
\end{equation}}
Therefore 
\begin{equation}
    \bra{x}e^{-\frac{i}{2}y\hat{ {\pi}}}e^{\frac{i}{2}y'\hat{ {\pi}}}e^{iq'\hat x} {X}_{{W}}(x',p')e^{\frac{i}{2}y'\hat{ {\pi}}}e^{-\frac{i}{2}y\hat{ {\pi}}}\ket{x}=e^{iq'x} {X}_{{W}}(x',p')\delta(y-y')
\end{equation}
Then it becomes easy to verify that \eqref{1.4} reduces to $ {X}_{{W}}(x,p)$ as required.

Conversely, substituting \eqref{1} in the RHS of \eqref{2}, we obtain
\begin{equation}
    (2\pi)^{-D}\int dq dy dx~ e^{\frac{i}{2}y\hat{ {\pi}}}e^{iq(\hat x-x)}\bra{x}e^{-\frac{i}{2}y\hat{ {\pi}}}\hat{ {X}} ~e^{-\frac{i}{2}y\hat{ {\pi}}}\ket{x}e^{\frac{i}{2}y\hat{ {\pi}}} \label{yo}
\end{equation}
We can write $\bra{x}e^{-\frac{i}{2}y\hat{ {\pi}}}\hat{ {X}} ~e^{-\frac{i}{2}y\hat{ {\pi}}}\ket{x}$ as
\begin{equation}
      U(x,x-y/2)\bra{x-y/2}\hat{  X}\ket{x+y/2}  U(x+y/2,x)
\end{equation}
Inserting a complete set of states afore \eqref{yo} we get
\begin{equation}
    \int dy dx~ \ket{x-y/2}\bra{x}\bra{x-y/2}\hat{ {X}}\ket{x+y/2}  U(x+y/2,x)e^{\frac{i}{2}y\hat{ {\pi}}}
\end{equation}
Inserting a complete set of states after the expression we get
\begin{equation}
    \int dy dx~ \ket{x-y/2}\bra{x-y/2}\hat{ {X}}\ket{x+y/2}\bra{x+y/2}
\end{equation}
It is evident that this is $\hat{  X}$, as required.
\end{proof}
We note that \eqref{1} can also be written as follows:
\begin{equation}
	  X_{{W}}(x,p)=\int dy~e^{ipy}  U(x , x-y/2)\bra{x-y/2}\hat{X}\ket{x+y/2}  U(x+y/2 , x)  \label{12}
\end{equation}
From this expression it is obvious that $X_{{W}}(x,p)$ is transformed under the gauge transformation $\Omega(x)$ as $X_{{W}}(x,p)\to \Omega(x) X_{{W}}(x,p) \Omega(x)^\dagger$, i.e. that it belongs to the adjoint representation.


Now consider the generalized Wigner transform of the composition of two operators using \eqref{2} twice in \eqref{1}:
\begin{align}
\begin{split}
    &(\hat{  X}\hat{ {Y}})_{{W}}(x,p)\\
    &=(2\pi)^{-4D}\int dw ~e^{ipw}\int dqdydzdk dq'dy'dz'dk'~e^{-i(yk+qz+y'k'+q'z')}\\
    &\quad\times \bra{x}e^{-\frac{i}{2}w\hat{ \pi}}e^{\frac i 2 y\hat{  {\pi}}}e^{iq\hat x}  X_{{W}}(z,k)e^{\frac i 2 y\hat{  {\pi}}}e^{\frac i 2 y'\hat{  {\pi}}}e^{iq'\hat x}  Y_{{W}}(z',k')e^{\frac i 2 y'\hat{  {\pi}}}e^{-\frac{i}{2}w\hat{ \pi}}\ket{x}\\
    &=(2\pi)^{-2D}\int dydk dy'dk'~e^{-iy(k-p)-iy'(k'-p)}\\
    &\quad\times  {  U(x , x-(y+y')/2)  U(x-(y+y')/2 , x-y'/2)  X_{{W}}(x-y'/2,k)   U(x-y'/2 , x+(y-y')/2)}\\
    &\quad\quad {  U(x+(y-y')/2, x+y/2)  Y_{{W}}(x+y/2,k')  U(x+y/2,  x+(y+y')/2)  U(x+(y+y')/2 , x)} \\
    &=X_{{W}}\bigstar Y_{{W}} \label{yoni}
\end{split}
\end{align}
where in the final line we have defined the generalized star product.

\subsection{Generalized Moyal product}
{Recall that $D=\partial-iA$. Now let us show that the generalized star product may be written as
\begin{align}
	\begin{split}
		&{  X}_W(x,p) \bigstar { {Y}}_W(x,p) =\\&\bigg(e^{\frac{i}{2}(\Rr\partial_{p_1}+\Rr\partial_{p_2})\Rr D_{x}}e^{-\frac{i}{2}\Rr\partial_{ {p_1}}\Rr D_{x}} {X}_W(x,p_1)e^{-\frac{i}{2}\Rr D_{x} {\Ll\partial_{p_1}}}\\\
		&e^{-\frac{i}{2} {\Rr\partial_{p_2}}\Rr{D}_{x}} {Y}_W(x,p_2)e^{-\frac{i}{2}\Ll\partial_{ {p_2}}\Rr{D}_{x}}e^{\frac{i}{2}(\Ll\partial_{p_1}+\Ll\partial_{p_2})\Rr{D}_{x}}\bigg)\times 1\,\bigg|_{\substack{p_1=p_2=p}} \label{yo2}
	\end{split}
\end{align}
We can trivially write $X_W(x,p)=(2\pi)^{-D}\int dkdy~ e^{-iy(k-p)}X_W(x,k)$ and $Y_W(x,p)=(2\pi)^{-D}\int dk'dy'~ e^{-iy'(k'-p)}Y_W(x,k')$. Then 
\begin{align}
	\begin{split}
		&{  X}_W(x,p) \bigstar { {Y}}_W(x,p)=\\
		&(2\pi)^{-2D}\int dydk dy'dk'~e^{-iy(k-p)-iy'(k'-p)}\\
		&e^{-\frac{y'+y}{2}\Rr{D}_{x}}e^{\frac{y}{2}\Rr{ D}_{x}} {X}_W(x,k')e^{\frac{y}{2}\Rr{D}_{x}}e^{\frac{y'}{2}\Rr{D}_{x}}\\&{Y}_W(x,k')e^{\frac{1}{2}y'\Rr{D}_{x}}e^{-\frac{y+y'}{2}\Rr{D}_{x}}\times 1 \label{ff?}
	\end{split}
\end{align}}
Now we use the following sequence of relations:
\begin{eqnarray}
&&	e^{-\frac{y'+y}{2}\Rr{D}_{x}}e^{\frac{y}{2}\Rr{ D}_{x}} {X}_W(x,k)e^{\frac{y}{2}\Rr{D}_{x}}e^{\frac{y'}{2}\Rr{D}_{x}}\nonumber\\&&{Y}_W(x,k')e^{\frac{1}{2}y'\Rr{D}_{x}}e^{-\frac{y+y'}{2}\Rr{D}_{x}}\times 1 \nonumber\\ && = 
e^{-\frac{y'+y}{2}\Rr{D}_{x}}e^{\frac{y}{2}\Rr{ D}_{x}} {X}_W(x,k)e^{\frac{y}{2}\Rr{D}_{x}}e^{\frac{y'}{2}\Rr{D}_{x}}\nonumber\\&&{Y}_W(x,k')e^{\frac{1}{2}y'\Rr{D}_{x}}U(x,x-\frac{y+y'}{2})\nonumber\\ && = 
e^{-\frac{y'+y}{2}\Rr{D}_{x}}e^{\frac{y}{2}\Rr{ D}_{x}} {X}_W(x,k)e^{\frac{y}{2}\Rr{D}_{x}}e^{\frac{y'}{2}\Rr{D}_{x}}\nonumber\\&&{Y}_W(x,k')U(x,x+y'/2)U(x+y'/2,x-\frac{y}{2}) \nonumber\\ && = 
e^{-\frac{y'+y}{2}\Rr{D}_{x}}e^{\frac{y}{2}\Rr{ D}_{x}} {X}_W(x,k)e^{\frac{y}{2}\Rr{D}_{x}}U(x,x+y'/2)\nonumber\\&&{Y}_W(x+y'/2,k')U(x+y'/2,x+y')U(x+y',x-\frac{y-y'}{2})
\nonumber\\ && = 
e^{-\frac{y'+y}{2}\Rr{D}_{x}}e^{\frac{y}{2}\Rr{ D}_{x}} {X}_W(x,k)U(x,x+y/2)U(x+y/2,x+(y+y')/2)\nonumber\\&&{Y}_W(x+(y'+y)/2,k')U(x+(y'+y)/2,x+y'+y/2)U(x+y'+y/2,x+\frac{y'}{2}) \nonumber\\ && = 
e^{-\frac{y'+y}{2}\Rr{D}_{x}}U(x,x+y/2) {X}_W(x+y/2,k)U(x+y/2,x+y)U(x+y,x+y+y'/2)\nonumber\\&&{Y}_W(x+y+y'/2,k')U(x+y+y'/2,x+y'+y)U(x+y'+y,x+\frac{y'+y}{2})  \nonumber\\ && =
U(x,x-(y+y')/2)U(x-(y+y')/2,x-y'/2) {X}_W(x-y'/2,k)\nonumber\\&&U(x-y'/2,x+y/2-y'/2)U(x+y/2-y'/2,x+y/2)\nonumber\\&&{Y}_W(x+y/2,k')U(x+y/2,x+y'/2+y/2)U(x+y'/2+y/2,x) 
\end{eqnarray}

One can also derive the following representation
{\begin{align}
	\begin{split}
		&{  X}_W(x,p) \bigstar { {Y}}_W(x,p)=\\
		&(2\pi)^{-2D}\int dydk dy'dk'~e^{-iy(k-p)-iy'(k'-p)}\\
		&e^{-\frac{y'+y}{2}\Rr{D}_{x}}e^{\frac{y}{2}\Rr{ D}_{x}}e^{\frac{y'}{2}\Rr{ D}_{x}}\Bigl(e^{-\frac{y'}{2}\Rr{\cal D}_{x}} {X}_W(x,k)\Bigr)e^{-\frac{y'}{2}\Rr{D}_{x}}e^{\frac{y}{2}\Rr{D}_{x}}e^{\frac{y'}{2}\Rr{D}_{x}}e^{-\frac{1}{2}y\Rr{D}_{x}}\\&\Bigl(e^{\frac{y}{2}\Rr{\cal D}_{x}}{Y}_W(x,k')\Bigr)e^{\frac{1}{2}y\Rr{D}_{x}}e^{\frac{1}{2}y'\Rr{D}_{x}}e^{-\frac{y+y'}{2}\Rr{D}_{x}}\times 1, \label{ff?}
	\end{split}
\end{align}}
where $\cal D$ is the covariant derivative in Adjoint representation. 

{Another useful representation is
\begin{align}
	\begin{split}
		&{  X}_W(x,p) \bigstar { {Y}}_W(x,p)=\\
		&(2\pi)^{-2D}\int dydk dy'dk'~e^{-iy(k-p)-iy'(k'-p)}\\
		&W(x,y,y')\Bigl(e^{\frac{(y+y')}{2}\Rr{\mathcal{D}}_x}e^{-y'\Rr{\mathcal{D}}_x}e^{-\frac{y}{2}\Rr{\cal D}_{x}}{X}_W(x,k)\Bigr)\Bigl(e^{\frac{(y+y')}{2}\Rr{\mathcal{D}}_x}e^{-\frac{y'}{2}\Rr{\mathcal{D}}_x}{Y}_W(x,k')\Bigr)
	\end{split}
\end{align}
where $W(x,y,y')$ is the following triangular Wilson loop:
\begin{align}
\begin{split}
    W(x,y,y')=&U(x,x-(y+y')/2)U(x-(y+y')/2,x+y/2-y'/2)\\
    &U(x+y/2-y'/2,x+y'/2+y/2)U(x+y'/2+y/2,x) 
\end{split}
\end{align}
}

\section{Noether procedure in terms of Wigner-Weyl calculus for axial {and vector} currents}
\label{Sect4}

\subsection{Partition function and the action in terms of Wigner-Weyl calculus}

\mzc{We consider the $3+1$ D theory after Wick rotation to the Euclidean space - time. All external fields in the original theory do not depend on time, which allows us to deal with the technique of equilibrium theory. 
Among other external fields we distinguish the non-Abelian gauge field $A_\mu(x)$ (with gauge group $G$). It is coupled minimally to fermions from the fundamental representation.} The partition function of the theory is given above in Eq. (\ref{pf0}). The notions of covariant derivative  $D_\mu\equiv\partial_\mu-iA_\mu$ and Dirac operator $Q(x,-iD)$ are  defined in Sect. \ref{SectMain1}, Sect \ref{SectMain2}. 
The `inhomogeneity' in this model is contained in the explicit $x$ dependence in $Q(x,-iD)$.

In operator notation, the action can be written as 
\begin{equation}
    S=\text{tr}_{ D}\text{tr}_{ G}\text{tr}_{H}\left(\hat Q\hat\rho\right)
\end{equation}
where $\hat Q:=Q(\hat x,\hat \pi)$ (recall $\hat \pi:=\hat p-A(\hat x)$) and $\bra{x}\hat{\rho}\ket{y}:=\psi(x)\bar\psi(y)$ and $\text{tr}_D$, $\text{tr}_G$ and $\text{tr}_H$ are traces w.r.t. the spinor indices \mz{(and the other internal indices, excluding the gauge ones)}, gauge indices and the Hilbert space respectively.

We note the following exact relationship:
\begin{equation}
    Q_{{W}}(x,p)\equiv(\hat Q)_{{W}}(x,p)=Q(x,p) \label{exact}
\end{equation} 
which one can verify by taking the Weyl transform of $Q(x,p)$ using \eqref{2}. This formula means that $Q_{{W}}$ is independent of the gauge field and acts as the identity matrix on the fundamental representation.

\subsection{Vector current}

Consider the following non-Abelian, local gauge transformation:
\begin{align}
\psi(x)\to e^{i\alpha(x)}\psi(x) \label{thisv} \\
\bar\psi(x)\to\bar\psi(x)e^{-i\alpha(x)} \label{this1v} 
\end{align}
where $\alpha(x)$ belongs to the Lie algebra.

The variation of the action under an infinitesimal version of \eqref{thisv} and \eqref{this1v} is
\begin{align}
    \delta S&=\text{tr}_{ D}\text{tr}_{ G}\text{tr}_{H}\left(\delta\hat Q\hat\rho\right)\label{fade}\\
    \delta \hat Q&=[\hat Q,i\alpha(\hat x)] \label{rundmcv}
\end{align}
Using the trace this can be written as
\begin{align}
    \delta S=-\text{tr}_{ D}\text{tr}_{ G}\text{tr}_{H}\left(i\alpha(\hat x)[\hat Q,\hat\rho]\right) \label{faderv}
\end{align}
Using the fact that $\text{tr}_{G}\text{tr}_{H}(\hat X\hat Y)=\text{tr}_{G}\text{tr}_{\Gamma}(X_WY_W)$ (where $\text{tr}_\Gamma\equiv\int d^4x\int(2\pi)^{-4}d^4p$ is the trace w.r.t. the phase space) we get
\begin{align}
    \delta S&={+}\text{tr}_{G}\int d^4x~\alpha(x)\Gamma(x)\\
    \Gamma(x)&={-}i\text{tr}_{D}\int\frac{d^4p}{(2\pi)^{4}}(Q\bigstar\rho_W-\rho_W\bigstar Q) \label{ug}
\end{align}
Using \eqref{yoni} (the definition of the star) and integrating over $p$ we get
\begin{align}
\begin{split}
    \Gamma(x)&={-}i\text{tr}_{D}\int(2\pi)^{-8}d^4yd^4kd^4k'e^{-iy(k-k')}U(x,x+y/2)\\
    &\left(Q(x+y/2,k)\rho_W(x+y/2,k')-\rho_W(x+y/2,k)Q(x+y/2,k')\right)
    U(x+y/2,x) \label{geov}
    \end{split}
    \end{align}
Using \eqref{yonez} we can write \eqref{geov} as
\begin{align}
   \Gamma(x)&={-}i\text{tr}_{D}\int(2\pi)^{-8}d^4yd^4kd^4k'e^{-iy(k-k')}e^{y{\mathcal{D}}/2}\left(Q(x,k)\rho_W(x,k')-\rho_W(x,k)Q(x,k')\right) 
\end{align}
Integrating by parts we get
\begin{align}
   \Gamma(x)&={-}i\text{tr}_{D}\int(2\pi)^{-8}d^4yd^4kd^4k'e^{-iy(k-k')}\left(e^{-i\partial_{k}{\mathcal{D}}/2}\left(Q(x,k)\rho_W(x,k')\right)-e^{i\partial_{k'}{\mathcal{D}}/2}\left(\rho_W(x,k)Q(x,k')\right)\right) \label{her}
\end{align}
Expanding in powers of the covariant derivative 
we get
\begin{align}
   \Gamma&= {+}2 i \text{tr}_{D}\int\frac{d^4p}{(2\pi)^4}\sum_{k=0}^\infty\frac{(i/2)^{2k+1}}{(2k+1)!} \Big[\Pi_{i=1...2k+1}{\mathcal{D}}_{\mu_i}\Big]\left(\rho_W\Big[\Pi_{i=1...2k+1}\partial_{p_{\mu_i}}\Big]Q\right)
\end{align}
and 
\begin{align}
	\Gamma&= {\mathcal{D}}_\mu J_\mu \\
	J_{\mu}&={-}\text{tr}_{D}\int\frac{d^4p}{(2\pi)^4}\sum_{k=0}^\infty\frac{(-1)^k }{4^k(2k+1)!} \Big[\Pi_{i=1...2k}{\mathcal{D}}_{\mu_i}\Big]\left(\rho_W\Big[\Pi_{i=1...2k}\partial_{p_{\mu_i}}\Big] \partial_{p_\mu} Q\right)
\end{align}
The expression for the vector current is simplified considerably in two important cases:
if 
\begin{align}
	\partial_{p_{\mu_1}}...\partial_{p_{\mu_n}}Q(x,p)=0\quad\text{for }n\geq 3
\end{align}
or if we can neglect 
\begin{align}
\frac{1}{4^k (2k+1!)}{\mathcal{D}}_{\mu_1}...{\mathcal{D}}_{\mu_{2k}}	\partial_{p_{\mu_1}}...\partial_{p_{\mu_{2k}}}\partial_{p_{\mu_{0}}}Q(x,p)\quad\text{for }k\geq 1 \label{ddQ}
\end{align}
compared to $\partial_{p_{\mu_{0}}}Q(x,p)$
because the inhomogeneity is sufficiently weak, i.e. the higher derivatives of the background fields entering the Dirac operator may be neglected\footnote{In order to better understand the nature of the second case mentioned above let us consider the situation in which $Q(x,p) \sim f(x) \, \frac{{\rm sin}\, (p a)}{a}$. Here the function $f(x)$ models the background fields while the parameter $a$ represents the scale specific to the Dirac operator (if the background fields were constant). Then  (\ref{ddQ}) gives $\frac{d^{2k}}{d^{2k} x}f(x) a^{2k} {\rm sin}\, (pa)$ which is to be compared with $f(x) {\rm sin}\, (pa)$. This way we arrive at the condition $\frac{d^{2k}}{d^{2k} x}f(x) a^{2k}  \ll f(x) $: i.e. that the function $f(x)$ varies slowly enough that its variation at the distance of the order of $a$ is to be neglected.}. 
In both these cases we arrive at
\begin{align}
    J_{\mu}&=-\text{tr}_{D}\int\frac{d^4p}{(2\pi)^4}\rho_W\partial_{p_\mu}Q \label{curv}
\end{align}
{We can come to the same formula in a different way. Namely, let us express the action as 
\begin{align}
	\delta S=-i\text{tr}_{ D}\text{tr}_{ G}\text{tr}_{H}\left(\hat\rho [\alpha(\hat x),\hat Q]\right) \label{faderv2}
\end{align}
Using the fact that $\text{tr}_{G}\text{tr}_{H}(\hat X\hat Y)=\text{tr}_{G}\text{tr}_{\Gamma}(X_WY_W)$ (where $\text{tr}_\Gamma\equiv\int d^4x\int(2\pi)^{-4}d^4p$ is the trace w.r.t. the phase space) we get
\begin{align}
	\delta S&=-i{\text{tr}_D}\text{tr}_{G}\int d^4x~\frac{d^4p}{(2\pi)^{4}}\rho_W(x,p)(\alpha(x)\bigstar Q(x,p)-Q(x,p) \bigstar \alpha(x)) \label{ug2}
\end{align}
Using \eqref{yoni} (the definition of the star) we get
\begin{align}
	\delta S&=-i{\text{tr}_D}\text{tr}_{G}\int d^4x~\frac{d^4p}{(2\pi)^{4}}\rho_W(x,p)\Bigl( (2\pi)^{-2D}\int dydk dy'dk'~e^{-iy(k-p)-iy'(k'-p)}\nonumber\\
	&\quad\times  {  U(x , x-(y+y')/2)  U(x-(y+y')/2 , x-y'/2)  \alpha(x-y'/2)   U(x-y'/2 , x+(y-y')/2)}\nonumber\\
	&\quad\quad {  U(x+(y-y')/2, x+y/2)  Q(x+y/2,k')  U(x+y/2,  x+(y+y')/2)  U(x+(y+y')/2 , x)} \nonumber\\ & \quad  - (\alpha(x) \leftrightarrow Q(x,p))\Bigr)\nonumber\\
	&=-i{\text{tr}_D}\text{tr}_{G}\int d^4x~\frac{d^4p}{(2\pi)^{4}}\rho_W(x,p)\Bigl( (2\pi)^{-D}\int  dy'dk'~e^{-iy'(k'-p)}\nonumber\\
	&\quad\times  {  U(x , x-y'/2)    \alpha(x-y'/2)    U(x-y'/2, x)  Q(x,k')  }   - (\alpha(x) \leftrightarrow Q(x,p))\Bigr)
	\nonumber\\
	&=-i{\text{tr}_D}\text{tr}_{G}\int d^4x~\frac{d^4p}{(2\pi)^{4}}\rho_W(x,p) (2\pi)^{-D}\int  dy'dk'~e^{-iy'(k'-p)}\nonumber\\
	&\quad\times \Bigl( { [e^{-y'{\mathcal{D}}/2}   \alpha(x) ]   Q(x,k')  }   - Q(x,k')  [e^{y'{\mathcal{D}}/2}   \alpha(x) ]\Bigr) \label{ug2}
\end{align}
Now let us assume that $\alpha(x)$ is varying smoothly, so that we can neglect the derivatives higher than the first ones. Then we arrive at 
\begin{align}
	\delta S&=-i{\text{tr}_D}\text{tr}_{G}\int d^4x~\alpha(x)\,{\mathcal D_\mu}\,\int \frac{d^4p}{(2\pi)^{4}}(2\pi)^{-D}\int  dy'dk'~e^{-iy'(k'-p)} \rho_W(x,p)    {y^{\prime\mu}}   Q(x,k')\nonumber\\ 
	&={+\text{tr}_D}\text{tr}_{G}\int d^4x~\alpha(x)\,{\mathcal D_{\mu}}\,\int \frac{d^4p}{(2\pi)^{4}}(2\pi)^{-D}\int  dy'dk'~{\partial_{k'_\mu}} e^{-iy'(k'-p)} \rho_W(x,p)       Q(x,k')  \nonumber\\ 
	&=-{\text{tr}_D}\text{tr}_{G}\int d^4x~\alpha(x)\,{\mathcal D_\mu}\,\int \frac{d^4p}{(2\pi)^{4}}  \rho_W(x,p)     {\partial_{p_\mu}}  Q(x,p)    \label{ug2}
\end{align}}
The expectation value of the current is
	\begin{equation}
		\langle J_{\mu}\rangle=-\text{tr}_{D}\int\frac{d^4p}{(2\pi)^4}G_W\partial_{p_\mu} Q \label{unrah}
	\end{equation}
where $G_W:=\langle\rho_W \rangle$. We also have $G_W=(\hat G)_W$ where $\hat G=\hat Q^{-1}$ -- i.e.
\begin{equation}
    \hat Q\hat G=1 \label{identity}
\end{equation}

\subsection{Axial current}

Now consider the following non-Abelian, local chiral transformation:
\begin{align}
	\psi(x)\to e^{i\alpha(x)\gamma^5}\psi(x) \label{this} \\
	\bar\psi(x)\to\bar\psi(x)e^{i\alpha(x)\gamma^5} \label{this1} 
\end{align}
where $\alpha(x)$ is in the Lie algebra.

The variation of the action under an infinitesimal version of \eqref{this} and \eqref{this1} is
\begin{align}
	\delta S&=\text{tr}_{ D}\text{tr}_{ G}\text{tr}_{H}\left(\delta\hat Q\hat\rho\right)\label{fade}\\
	\delta \hat Q&=\{\hat Q,i\alpha(\hat x)\gamma^5\} \label{rundmc}
\end{align}
Using the trace this can be written as
\begin{align}
	\delta S=\text{tr}_{ D}\text{tr}_{ G}\text{tr}_{H}\left(i\alpha(\hat x)\gamma^5\{\hat Q,\hat\rho\}\right) \label{fader}
\end{align}
Using the fact that $\text{tr}_{ G}\text{tr}_{H}(\hat X\hat Y)=\text{tr}_{G}\text{tr}_{\Gamma}(X_WY_W)$ (where $\text{tr}_\Gamma\equiv\int d^4x\int(2\pi)^{-4}d^4p$ is the trace w.r.t. the phase space) we get
\begin{align}
	\delta S&=\text{tr}_{G}\int d^4x~\alpha(x)\Gamma(x)\\
	\Gamma(x)&=i\text{tr}_{D}\gamma^5\int\frac{d^4p}{(2\pi)^{4}}(Q\bigstar\rho_W+\rho_W\bigstar Q) \label{ug}
\end{align}
Using \eqref{yoni} (the definition of the star) and integrating over $p$ we get
\begin{align}
	\begin{split}
		\Gamma(x)&=i\text{tr}_{D}\gamma^5\int(2\pi)^{-8}d^4yd^4kd^4k'e^{-iy(k-k')}U(x,x+y/2)\\
		&\left(Q(x+y/2,k)\rho_W(x+y/2,k')+\rho_W(x+y/2,k)Q(x+y/2,k')\right)
		U(x+y/2,x) \label{geo}
	\end{split}
\end{align}
Using \eqref{yonez} we can write \eqref{geo} as
\begin{align}
	\Gamma(x)&=i\text{tr}_{D}\gamma^5\int(2\pi)^{-8}d^4yd^4kd^4k'e^{-iy(k-k')}e^{y{\mathcal{D}}/2}\left(Q(x,k)\rho_W(x,k')+\rho_W(x,k)Q(x,k')\right) 
\end{align}
Integrating by parts we get
\begin{align}
	\Gamma(x)&=i\text{tr}_{D}\gamma^5\int(2\pi)^{-8}d^4yd^4kd^4k'e^{-iy(k-k')}\left(e^{-i\partial_{k}{\mathcal{D}}/2}\left(Q(x,k)\rho_W(x,k')\right)+e^{i\partial_{k'}{\mathcal{D}}/2}\left(\rho_W(x,k)Q(x,k')\right)\right) \label{her}
\end{align}
Expanding in powers of the covariant derivative and assuming that $Q(x,p)$ is such that 
\begin{align}
	\partial_{p_{\mu_1}}...\partial_{p_{\mu_n}}Q(x,p)=0\quad\text{for }n\geq 3
\end{align}
or 
\begin{align}
	\Big|\frac{1}{4^k (2k+1!)}{\mathcal{D}}_{\mu_1}...{\mathcal{D}}_{\mu_{2k}}	\partial_{p_{\mu_1}}...\partial_{p_{\mu_{2k}}}\partial_{p_{\mu_{0}}}Q(x,p) \Big| \ll \Big|\partial_{p_{\mu_{0}}}Q(x,p) \Big| \quad\text{for }k\geq 1 \label{ddQ}
\end{align}
we get
\begin{align}
	\Gamma&=i\text{tr}_{D}\gamma^5\int\frac{d^4p}{(2\pi)^4}\{Q,\rho_W\}+{\mathcal{D}}_\mu J^{(5)}_\mu \\
	J^{(5)}_{\mu}&=\frac{1}{2}\text{tr}_{D}\gamma^5\int\frac{d^4p}{(2\pi)^4}\left(\partial_{p_\mu}Q\rho_W-\rho_W\partial_{p_\mu}Q\right)
\end{align}
The same expression appears if we assume that function $\alpha(x)$ varies slowly.

{Let us suppose for a moment that  $\{\gamma^5,Q\}=0$ (i.e. classical chiral symmetry -- it is actually broken by regularization). Then} we get
\begin{align}
	\Gamma&={\mathcal{D}}_\mu J^{(5)}_\mu \label{sates}\\
	J^{(5)}_{\mu}&=-\text{tr}_{D}\gamma^5\int\frac{d^4p}{(2\pi)^4}\rho_W\partial_{p_\mu}Q \label{cur}
\end{align} 

When the equations of motion are satisfied, then $\delta S=0$ -- which, since $\alpha(x)$ is arbitrary, leads to:
\begin{align}
	\Gamma(x)=0\quad \text{(when EOMs are satisfied)} \label{satisfes}
\end{align}
Using \eqref{sates} we get the classical conservation law:
\begin{align}
	{\mathcal{D}}_{\mu}J^{(5)}_\mu=0\quad \text{(when EOMs are satisfied)}
\end{align}
{This equation is broken due to the chiral anomaly, which becomes obvious when an ultraviolet regularization that does not respect chiral symmetry is applied.}
The expectation value of the current is
\begin{equation}
	\langle J^{(5)}_{\mu}\rangle=-\frac{1}{2}\text{tr}_{D}\int\frac{d^4p}{(2\pi)^4}G_W\partial_{p_\mu}[Q,\gamma^5]  \label{unrah}
\end{equation}
\mz{\it Recall that if the Dirac operator contains extra indices (in addition to the Dirac spinor index and the index of the given gauge group of the field $A$) we assume that the symbol $\text{tr}_D$ contains both these indices and the spinor ones. }

\section{Iterative solution of the Generalized Groenewold equation}
\label{Sect5}

The Wigner transform of the identity operator is $1$. Taking the Wigner transform of \eqref{identity} and using \eqref{yoni} and \eqref{exact} then gives
\begin{align}
\begin{split}
    1=&(2\pi)^{-2D}\int dydk dy'dk'~e^{-iy(k-p)-iy'(k'-p)}Q(x-y'/2,k)\\
    &  U(x , x-(y+y')/2)  U(x-(y+y')/2 , x-y'/2)     U(x-y'/2 , x+(y-y')/2)\\
    &U(x+(y-y')/2, x+y/2)  G_{{W}}(x+y/2,k')  U(x+y/2,  x+(y+y')/2)  U(x+(y+y')/2 , x) \label{id1}
\end{split}
\end{align} 
To distinguish between the $x$ dependence arising from the inhomogeneity vs from the gauge field we introduce the function $G_{{W}}(x,p,z)$ (the $z$ dependence arises from the gauge field while the $x$ dependence from the inhomogeneity) which is the solution of 
\begin{align}
\begin{split}
    1=&(2\pi)^{-2D}\int dydk dy'dk'~e^{-iy(k-p)-iy'(k'-p)}Q(x-y'/2,k)\\
    &  U(z , z-(y+y')/2)  U(z-(y+y')/2 , z-y'/2)     U(z-y'/2 , z+(y-y')/2)\\
    &U(z+(y-y')/2, z+y/2)  G_{{W}}(x+y/2,k',z+y/2)  U(z+y/2,  z+(y+y')/2)  U(z+(y+y')/2 , z) \label{id1''}
\end{split}
\end{align}
Comparing with \eqref{id1} it is easy to see that $G_{{W}}(x,p)=G_{{W}}(x,p,x)$. Note that $Q(x,p)$ has no $z$ dependence.
Using  (\ref{ff?}) we rewrite  (\ref{id1''}) as
\begin{align}
\begin{split}
    1=&(2\pi)^{-2D}\int dydk dy'dk'~e^{-iy(k-p)-iy'(k'-p)}Q(x-y'/2,k)\\
    &  {e^{-\frac{1}{2}(y+y')D_z}e^{yD_z}e^{\frac{y'}{2}D_z}G_{{W}}(x+y/2,k',z)e^{\frac{y'}{2}{D}_z}e^{-\frac{1}{2}(y+y'){D}_z}  }\times 1 \label{id1'}
\end{split}
\end{align}
where $D_z$ is the gauge covariant derivative w.r.t. $z$ { -- it acts on the right -- and $1$ means the function of $z$ that is equal to $1$ and thus does not depend on $z$ at all.}
We will expand  (\ref{id1'}) in powers of $D_z$ and expand $G_{{W}}$ as
\begin{equation}
    G_{{W}}(x,p,z)=\sum_{n\geq 0} G^{(n)}(x,p,z)
\end{equation}
where $G^{(n)}(x,p,z)$ contains $n$ powers of $D_z$. We will suppose that $G^{(0)}(x,p,z)\equiv G^{(0)}(x,p)$ is independent of $A(z)$, so we can take it to be proportional to the identity matrix in gauge space. 

In what follows we will make use of the fact that the triangular Wilson loop 
\begin{align}
\begin{split}
&U(z,z-(y+y')/2) U(z-(y+y')/2,z+(y-y')/2)\\&U(z+(y-y')/2,z+(y+y')/2) U(z+(y+y')/2,z) \\
&=e^{-\frac{1}{2}(y+y')D_z}e^{yD_z}e^{y'D_z}e^{-\frac{1}{2}(y+y')D_z}\times 1
\end{split}
\end{align}
can be expanded in powers of $D_z$ as
\begin{align}
\begin{split}
&1-\frac{i}{2}y^\mu y'^\nu F_{\mu\nu}(z)-\frac{i}{12}(y^\alpha-y'^\alpha)y^\mu y'^\nu \mz{\cal D}_\alpha F_{\mu\nu}-\frac{i}{48}(y^\alpha y^\beta+y'^\alpha y'^\beta) y^\mu y'^\nu \mz{\cal D}_{\alpha}\mz{\cal D}_\beta F_{\mu\nu}(z)\\
   &-\frac{1}{8}y^\mu y'^\nu y^\alpha y'^\beta F_{\mu\nu}(z)F_{\alpha\beta}(z)+\mathcal O(D_z^5)
\end{split}
\end{align}
Now, to lowest order in $D_z$, \eqref{id1'} becomes
\begin{equation}
    Q\star G^{(0)}=1 \label{bumb}
\end{equation}
where $\star = e^{\frac{i}{2}(\overleftarrow{\partial}_x\overrightarrow{\partial}_p-\overleftarrow{\partial}_p\overrightarrow{\partial}_x)}$ is the ordinary star product.
Expanding \eqref{id1'} to order one in $D_z$ proves that $G^{(1)}=0$. 
Expanding to order two gives
\begin{equation}
    0=Q\star G^{(2)}+\frac{i}{2}\partial_{p_\mu}Q\star \partial_{p_\nu}G^{(0)}F_{\mu\nu}(z)
\end{equation}
where we've made use of the fact that 
$[D_\mu,D_\nu]=-iF_{\mu\nu}$
-- whose solution is
\begin{align}
    G^{(2)}(x,p,z)&=-\frac{i}{2}G^{(0)}(x,p)\star\partial_{p_{\mu}}Q(x,p)\star \partial_{p_{\nu}}G^{(0)}(x,p)F_{\mu\nu}(z)\label{G2}
\end{align}
We will define
\begin{align}
   \Gamma_{\mu\nu}&:=-\frac{i}{4}G^{(0)}\star\partial_{p_{[\mu}}Q\star \partial_{p_{\nu]}}G^{(0)} \label{imus}
\end{align}
So 
\begin{equation}
    G_{{W}}=G^{(0)}+\Gamma_{\mu\nu}F_{\mu\nu}+O(D^3) \label{im}
\end{equation}

\section{Axial current in Pauli-Villars regularization and non-Abelian CSE}
\label{Sect6}

\subsection{General considerations}

\mzc{In this paper we use the  Pauli-Villars regularization (for the detailed definitions see above Sect. \ref{SectMain3}). The grand thermodynamic potential is usually defined as $\Omega = - T \,{\rm log}\, Z$ (where $T$) is temperature. In the given regularization this expression is modified as 
	\begin{equation}
	\Omega = - T\,	\sum_{i} C_i \,{\rm log} \, Z(M_i),	
	\end{equation}}
Here $Z(M_i)$ is calculated as  (\ref{pf0}) with $Q$ substituted by 
	$Q_{(M_i)}(i\partial_p, p - A(i\partial_p))$ such that the nonzero parameter $M_i$ provides the appearance of a gap (i.e. a mass for the fermion). \mzc{The coefficients $C_i$ obey the conditions of Eq. (\ref{cond}).  
	These conditions provide the cancellation of ultraviolet divergencies. All macroscopic quantities may be calculated as combinations of $\Omega$ and its derivatives with respect to various parameters.} Following the previous steps we come, then, to the conclusion that in the regularized theory the axial current is
	\begin{equation}
		J^{(5)a}_\mu \equiv -\frac{1}{2}\sum_iC_i\int_{\mathcal M} \frac{dp}{(2\pi)^4}\text{tr}_D\text{tr}_G\left( G_{(M_i),W}\partial_{p_\mu}[Q_{(M_i)},t^a\gamma^5]\right)
	\end{equation}
\mz{Here $t^a$ is the generator of the gauge group under consideration.}
We substitute here  (\ref{im}) and obtain 
\begin{align}
	\begin{split}
		J^{(5)a}_\mu&=S_{\mu}(x)\text{tr}_{G} t^a+M_{\mu\alpha\beta}(x)\text{tr}_{G}(t^a F_{\alpha\beta}) \label{bruha}
	\end{split} \\
	S_{\mu}(x)&:=\frac{1}{2}\sum_iC_i\int \frac{dp}{(2\pi)^4}\text{tr}_{D}\left([\gamma^5, \partial_{p^\mu}Q]G^{(0)}\right)\\
	M_{\mu\alpha\beta}(x)&:=\frac{1}{2}\sum_iC_i\int \frac{dp}{(2\pi)^4}\text{tr}_{D}\left([t^a\gamma^5, \partial_{p^\mu}Q]\Gamma_{\alpha\beta}\right)
\end{align}
We use  (\ref{G2}) and obtain
\begin{align}
	M_{\mu\alpha\beta}(x)&:=-\frac{i}{8}\sum_iC_i\int \frac{dp}{(2\pi)^4}\text{tr}_{D}\left([\gamma^5, \partial_{p^\mu}Q]G^{(0)}\star\partial_{p^{[\alpha}}Q\star \partial_{p^{\beta]}}G^{(0)}\right)\label{M123}
\end{align}
We define the conductivity of the non-Abelian chiral separation effect as the response of the spatial components of the axial current to both the external magnetic field strength and to the chemical potential:
\begin{equation}
	\frac{\partial}{\partial \mu} J^{CSE}_{i} \equiv \frac{\partial}{\partial \mu} M_{ijk}\epsilon_{jkl} B_l
\end{equation}
Chemical potential appears as the shift of $p_4$ : $ i p_4 \to ip_4 + \mu$. Therefore, 
\begin{align}
	\frac{\partial }{\partial \mu}M_{ijk}(x)&=-\frac{1}{8}\sum_iC_i\int \frac{dp}{(2\pi)^4}\frac{\partial}{\partial p_4}\text{tr}_{D}\left([\gamma^5, \partial_{p^i}Q]G^{(0)}\star\partial_{p^{[j}}Q\star \partial_{p^{k]}}G^{(0)}\right)\label{M123mu}
\end{align}
Due to the presence of poles in the expression standing inside the integral, the integral is not zero, but it is reduced to integrals over the hyperplanes $\Sigma_{p_4}$ for  $p_4 = \pm \epsilon, \pm \infty$, $\epsilon \to 0$: 
\begin{align}
	\frac{\partial }{\partial \mu}M_{ijk}(x)&=-\frac{1}{8}\sum_iC_i\sum_{p_4 = \pm \epsilon, \pm \infty}\int_{\Sigma_{p_4}} \frac{dp_1 dp_2 dp_3}{(2\pi)^4}\text{tr}_{D}\left([\gamma^5, \partial_{p^i}Q]G^{(0)}\star\partial_{p^{[j}}Q\star \partial_{p^{k]}}G^{(0)}\right)\label{M123mu}
\end{align}
At this point we use the following  properties of the given system:
1) it is chiral close to the Fermi surface, i.e. at low energies (this does not concern the Pauli-Villars massive fields);
2) it is chiral in the ultraviolet, i.e. at large momenta (this concerns also the Pauli-Villars fields). Using  the properties of  (\ref{cond}) we come to conclusion that the contributions of the integrals at $\Sigma_{\pm \infty}$ cancel each other, the Pauli-Villars regulators disappear from the above expression, and we are left with the itegral over $\Sigma_{\pm \epsilon}$ for the original \mzo{massless} field only. The latter integrals can be reduced to integrals over \mzo{the  hypersurfaces $p_4 = \pm \epsilon$} that enclose the poles of the expression standing inside the integral: 
\begin{equation}
	\frac{\partial}{\partial \mu}\langle J^{(CSE)}_i\rangle = \sigma_{CSE} B_i.
\end{equation}
Here \mzc{$B_i=\frac{1}{2}\epsilon_{ijk}F_{jk}$ is the colormagnetic field}, averaging on the left hand side is over the whole system volume, while 
\begin{equation}
	\sigma_{CSE} = \frac{{ N}_3}{2\pi^2} \label{sigmaCSE}
\end{equation}
and
\mzo{\begin{eqnarray}
	N_3&=&\mz{-}\frac{1}{V \,48\pi^2}\sum_{p_4 = \pm \epsilon}\int d^3 x\int_{\Sigma_{p_4}} \text{tr}_{D}\left(G^{(0)}\star dQ\star \wedge dG^{(0)}\star\wedge dQ \gamma^5\right)\label{N3CSE}
\end{eqnarray}}
\mzb{The expression in the  above expression is a topological invariant due to the presence of integration over the three - dimensional momenta $\vec{p}$ and coordinates $\vec{x}$. There exists also the representation of Eq. (\ref{N3CSE}) in the form of an integral over the compact hypersurface \footnote{\mzb{In particular, we can represent \begin{equation} 
			N_3	={-}\frac{1}{V \,48\pi^2}\int d^3 x\int_{\Sigma_0} \text{tr}_{D}\left(\tilde{G}^{(0)}\star d\tilde{Q}\star \wedge d\tilde{G}^{(0)}\star\wedge d\tilde{Q} \gamma^5\right) 
			\nonumber
		\end{equation}
		In transition from Eq. (\ref{N3CSE}) to this expression we perform a smooth modification of Dirac operator: $Q((\vec{p},\pm \epsilon), x) \to Q((\vec{p},\pm \omega(\vec{p})),{x}) \equiv \tilde{Q}(\vec{p},\vec{x})$. Here $\omega(\vec{p})$ is a function, which describes the dependence of $p_4$ on $\vec{p}$ on the hypersurface $\Sigma_0$. The latter hypersurface in momentum space for any $x$ embraces the singularities of the expression standing inside the integral. For the case of a homogeneous system these singularities appear along the Fermi surface. In the inhomogeneous case the geometrical place in momentum space of such singularities depends on $x$ and generalizes the notion of the Fermi surface.
	In turn, in transition from Eq. (\ref{N3CSE}) we deform also the Green function as $G^{(0)}(\vec{p},\pm \epsilon,\vec{x}) \to \tilde{G}^{(0)}(\vec{p},\vec{x})$, where the latter function obeys the {\it transformed} Groenewold equation 
		$$
		\tilde{Q}^{}(\vec{p},\vec{x})e^{\frac{i}{2}(\overleftarrow{\partial}_{\vec{x}}\overrightarrow{\partial}_{\vec{p}} - \overleftarrow{\partial}_{\vec{p}}\overrightarrow{\partial}_{\vec{x}})} \tilde{G}^{(0)}(\vec{p},\vec{x})
		$$
		The given smooth modification $Q \to \tilde{Q}$ and $G^{(0)}\to \tilde{G}^{(0)}$ does not result in a jump of the value of $N_3$ because defined this way it is the topological invariant.   
		}}  embracing the singularities of expression standing in the integral in Eq. (\ref{N3CSE}). }

\subsection{Illustravite examples}

\subsubsection{Massless Dirac fermions in the presence of background $U(1)$ gauge field and background vierbein}
\label{case1}

\mzc{For the purpose of illustration let us consider first the model of massless Dirac fermions $\psi$ in the presence of external $SU(N)$ gauge field $A$ and background fields: the $U(1)$ gauge field $C$ and vierbein $e_a^\mu$.

	Here we consider the systems with Dirac operator $\hat{Q}$ of the form: 
\begin{align}
	\begin{split}
		\hat{Q} &=  \gamma^a\{|e| e^\mu_{a}, (\hat p_\mu - A_\mu - { C}_\mu)\}/2 - m |e| \label{QB}
	\end{split}
\end{align}
The vierbein $e_a^\mu(x)$ depends on coordinates, $|e| = {\rm det}^{-1}[e^\mu_a]$ is the inverse determinant of matrix with components $e_a^\mu(x)$.
This is the general form of the action for Dirac fermions in the presence of Riemann - Cartan gravity (with vanishing spin connection but with nonzero torsion) and $U(1)$ gauge field $B$. For the Pauli - Villars regulators the values of $m$ are finite and are assumed to be large, while the original Dirac operator contains vanishing mass $m=0$.
Consider the case when the background fields are constant, and the Wigner transform of $\hat Q$ (at $A=0$) can be calculated and decomposed in terms of the gamma matrices as
	\begin{align}
		Q(p)=&\gamma^a |e|e^\mu_{a} (p_\mu - {C}_\mu) - m |e|\nonumber\\
		=&-\tilde{m}+\slashed \ell\nonumber\\
		\text{where}\quad\ell_a=&|e|e^{b\mu} \left(\delta_{ab}\left(p_\mu-C_\mu\right)\right), \quad \tilde{m} = m |e|
	\end{align}
	The inverse is therefore
	\begin{align}
		G^{(0)}(p)=Q^{-1}(p)&=c+\slashed r\nonumber\\
		c&=X^{-1}\tilde{m}\nonumber\\
		r_a&=X^{-1}\ell_a\nonumber\\
		\text{where}\quad X&=\tilde{m}^2+ \ell^2
	\end{align}
The conductivity of non - Abelian chiral separation effect is given by Eq. (\ref{sigmaCSE}) with $N_3$ of Eq. (\ref{N3CSE}) given by $N_3 = 1$ if the vierbein $e_a^\mu$ is nondegenerate  For the derivation see  \cite{ZA2023}. The calculation of Eq. (\ref{N3CSE}) is reduced to that of \cite{ZA2023} when matrix $e^\mu_a$ is continuously deformed to a unit matrix. If the number of Dirac fermions is multiplied and equals $n$, then $N_3=n$. Since the value of $N_3$ is a topological invariant, its value is not changed when the background fields $C$ and $e$ are slowly - varying. }

\subsubsection{$N_f$ Dirac fermions interacting with dynamical $SU(N_c)$ gauge fields }
\label{case2}
\mzc{Let us now consider the more involved case, when $C\in SU(N_c)$ is dynamical gauge field while the number of the flavors of the Dirac fermions experienced the background field $A$ and the dynamical field $C$ is $N_f$ (that is the total number of Dirac fermions is $N\times N_c \times N_f$). We consider the case, when $N \times N_f > 11 N_c/2$, in which  the theory does not possess chiral symmetry breaking \cite{Appelquist_1996}. (Say, we can take $N_c = 2$, $N = 3$, $N_f = 4$.) We also assume weak coupling of fermions to $C$.  In this situation again the topological invariant can be calculated using Eq. (\ref{N3CSE}), where we insert the interacting Green function instead of $G^{(0)}$, and its inverse instead of $Q$. Again, using technique of \cite{ZA2023} we obtain $N_3 = N_c \times N_f$.  }

\section{Vector current and non-Abelian QHE  }
\label{Sect7}
\subsection{Non-Abelian QHE in $3+1$ D systems}

{
For the vector current the above procedure may be modified accordingly, and we obtain instead of  (\ref{bruha}):
\begin{align}
	\begin{split}
		J^{}_\mu&=S^v_{\mu}(x)+M^v_{\mu\alpha\beta}(x) F_{\alpha\beta} \label{bruhv0}
	\end{split} 
\end{align}
with
\begin{align}
	M^{(v)}_{\mu\alpha\beta}(x)&:=\frac{i}{4}\sum_iC_i\int \frac{dp}{(2\pi)^4}\text{tr}_{D}\left( \partial_{p^\mu}QG^{(0)}\star\partial_{p^{[\alpha}}Q\star \partial_{p^{\beta]}}G^{(0)}\right)\nonumber\\  S^v_{\mu}(x)&=\sum_iC_i\int \frac{dp}{(2\pi)^4}\text{tr}_{D}\left( \partial_{p^\mu}Q G^{(0)}\right)
\end{align}
According to the Bloch theorem \cite{ZZ2019bloch} the first term in  (\ref{bruhv0}) for $\mu\ne 0$ vanishes because it is averaged over the spatial volume (total electric current vanishes in equilibrium), and we are left with the response of spatial current to the external field strength:
\begin{align}
	\begin{split}
		J^{}_\mu&= \epsilon_{\nu\mu\alpha\beta}M^{*(v)\nu}(x) F_{\alpha\beta} \label{bruhv}
	\end{split} 
\end{align}
with
\begin{align}
	M^{*(v)\nu}(x)&:=\frac{i}{2\times 3!}\epsilon^{\nu\mu\alpha\beta}\sum_iC_i\int \frac{dp}{(2\pi)^4}\text{tr}_{D}\left( \partial_{p^\mu}QG^{(0)}\star\partial_{p^{\alpha}}Q\star \partial_{p^{\beta}}G^{(0)}\right)
\end{align}
Let us consider the case of external field varying slowly. Response of spatial components of vector current (averaged over the system volume) to non-Abelian {\it electric} field strength (with nonzero $F_{4i} = - i E_i$ components) receives the form 
\begin{align}
	\begin{split}
		\bar{J}^{v,QHE}_i&=- \frac{1}{4\pi^2}\epsilon_{ijk}\tilde{M}^{*(v)j} E_{k} \label{QHEAG3}
	\end{split} 
\end{align}
with 
\begin{align}
	\begin{split}
		\tilde{M}^{*(v)i}&= 2i{M}^{*(v)i} = -\frac{1}{V \,24\pi^2}\sum_{i}C_i\int d^3 x  \wedge  dp^i  \wedge \text{tr}_{D}\left(G^{(0)}\star dQ\star \wedge dG^{(0)}\star\wedge dQ \right) 
	\end{split}\label{M3}
\end{align}
\mzc{In this expression $d^3x \equiv dx^1\wedge dx^2\wedge dx^3$, the integral is over the product of three - dimensional spatial coordinate space and four - dimensional momentum space, the sum over Pauli-Villars regulators is important in order to cancel the ultraviolet divergencies. In particular, this way the divergent contribution of the Dirac sea is cancelled.}
\mzc{At the same time vector current is an observable quantity, and its value cannot depend on the regularization. Below we will  consider particular examples of the models that possess non - Abelian QHE. We will see that in all considered cases the meaning of regulators in Eq. (\ref{M3}) is only to subtract the divergent contribution of the Dirac sea (if any). We, therefore, are able to simplify the above expression:
\begin{align}
	\begin{split}
		\tilde{M}^{*(v)i}&= 2i{M}^{*(v)i} = -\frac{1}{V \,24\pi^2}\Bigl[\int d^3 x \wedge  dp^i  \wedge \text{tr}_{D}\left(G^{(0)}\star dQ\star \wedge dG^{(0)}\star\wedge dQ \right) \Bigr]_{reg}
	\end{split}\label{M3c}
\end{align}
Here subscript "reg" means that the divergent contribution of the Dirac sea is to be subtracted from the considered quantity. This may particularly be achieved if we add the sum over the Pauli - Villars regulators as in Eq. (\ref{M3c}). }

{For the density of non-Abelian charge we obtain in a similar way the response to the external {\it magnetic} non-Abelian field $B_i$:
\begin{align}
	\begin{split}
		\bar{\rho}^{v}&=i J_4= \frac{1}{4\pi^2}\tilde{M}^{*(v)i} B_{i} \label{QHEAGM}
	\end{split} 
\end{align}
In addition to Eqs. (\ref{QHEAGM}) and (\ref{QHEAG3}) one can derive the  response of the non-Abelian current to the  (magnetic) field strength:
\begin{align}
	\begin{split}
		\bar{J}^{v(B)}_i&= \frac{1}{4\pi^2}\tilde{M}^{*(v)4} B_{i} \label{QHEAGCME}
	\end{split} 
\end{align}
Notice that $\tilde{M}^{*(v)4}$ is the same quantity that enters the expression for ordinary electric current. Due to the Bloch theorem, the latter is absent in equilibrium. Therefore, $\tilde{M}^{*(v)4}$ should vanish as well as the current of  (\ref{QHEAGCME}). }

{It is worth mentioning that \mz{if the fermions under consideration are electrically charged, the vector $\tilde{M}^{*(v)i}$ is related to the intrinsic magnetic moment ${\cal M}$ of the fermion system (see Appendix \ref{Magnetic} and\footnote{\mz{Notice that in \cite{Z2024CSE}) there is a mistake in the sign of the magnetic moment.}} Sect. 3 of \cite{Z2024CSE})}:
\mz{\begin{equation}
	\frac{1}{4\pi^2} 	\tilde{M}^{*(v)i} =\frac{\partial}{\partial \mu} {\cal M}_i 
\end{equation}} 
Therefore, we have
\begin{align}
	\begin{split}
		\vec{J}^{v,QHE}&= \mz{ \Bigl[\vec{E}\times \frac{\partial}{\partial \mu} \vec{\cal M}\Bigr]}\\
		{\rho}^{v,QHE}&= \mz{ \Bigl(\vec{B} \frac{\partial}{\partial \mu} \vec{\cal M}\Bigr)}  \label{QHEAGV}
	\end{split} 
\end{align} }

\subsection{Non-Abelian QHE in $2+1$ D }
Let us consider the $2+1$D gapped system. Again, as in $3+1$D, for the vector current we obtain as in  (\ref{QHEAG3}):
\begin{align}
	\begin{split}
		J^{v}_\mu&=S^v_{\mu}(x)+M^v_{\mu\alpha\beta}(x) F_{\alpha\beta} \label{bruhv2}
	\end{split} 
\end{align}
with
\begin{align}
	M^v_{\mu\alpha\beta}(x)&:=\frac{i}{4}\sum_iC_i\int \frac{dp}{(2\pi)^3}\text{tr}_{D}\left( \partial_{p^\mu}QG^{(0)}\star\partial_{p^{[\alpha}}Q\star \partial_{p^{\beta]}}G^{(0)}\right)\nonumber\\  S^v_{\mu}(x)&=\sum_iC_i\int \frac{dp}{(2\pi)^3}\text{tr}_{D}\left( \partial_{p^\mu}Q G^{(0)}\right)\\
\end{align}
Response of the vector current (spatial components averaged over the system volume) to the non-Abelian {\it electric} field strength is (for the case of external field varying slowly)
\begin{align}
	\begin{split}
		\bar{J}^{v,QHE}_i&= \frac{1}{2\pi}\epsilon_{ij}{M}^v_3 E_{j} \label{QHEAG}
	\end{split} 
\end{align}
with 
\begin{align}
	\begin{split}
		{M}^v_3&=-\frac{1}{S \,24\pi^2}\sum_i C_i \int d x^1 \wedge dx^2 \wedge  \text{tr}_{D}\left(G^{(0)}\star dQ\star \wedge dG^{(0)}\star\wedge dQ \right) 
	\end{split}\label{M3}
\end{align}
\mzc{Here $S$ is the area of the sample assumed to be large. Integration is over the product of two - dimensional spatial part of coordinate space and three - dimensional momentum space. As above, in the case of the four - dimensional system, the only meaning of the regulator fields is to subtract the divergent unphysical contribution (if any) of the Dirac sea. That's why we formulate Eq. (\ref{M3}) as  
\begin{align}
	\begin{split}
		{M}^v_3&=-\frac{1}{S \,24\pi^2}\Bigl[\int d^2 x\wedge  \text{tr}_{D}\left(G^{(0)}\star dQ\star \wedge dG^{(0)}\star\wedge dQ \right) \Bigr]_{reg}
	\end{split}\label{M3c}
\end{align} 
where $d^2x \equiv dx^1\wedge dx^2$, the subscript "reg" means the above mentioned subtraction of an unphysical contribution. One can see that in our approach this subtraction appears authomatically if one uses the Pauli - Villars regularization. However, the same result, of course, will be valid in any other regularization. In particular, in lattice regularization there is the peculiar mechamism of the subtraction of the Dirac see contributions. For the details in the applications to graphene see \cite{hatsugai2006topological}.} 

The response of the non-Abelian charge density to external {\it magnetic} field strength $B$ is
\begin{align}
	\begin{split}
		\bar{\rho}^{v}&= \frac{1}{2\pi} {M}^v_3 B \label{QHEAG}
	\end{split} 
\end{align}
In a similar way to that of Appendix \ref{Magnetic} one can prove that the magnetic moment $\cal M$ of the $2D$ system is related to $M^v_3$ as 
\begin{equation}
	\frac{\partial{\cal M}}{\partial \mu} = \mz{ \frac{1}{2\pi} M^v_3}
\end{equation}
Therefore, we represent the Hall current and the (non- Abelian) charge density as 
\begin{align}
	\begin{split}
		\vec{J}^{v,QHE}&= \mz{ ^*\vec{E}  \frac{\partial}{\partial \mu} {\cal M}}\\
		{\rho}^{v,QHE}&= \mz{ {B} \frac{\partial}{\partial \mu} {\cal M}},  \label{QHEAG2D}
	\end{split} 
\end{align} 
where $^*$ is the $2D$ Hodge duality operation.

\subsection{Illustrative examples}

\label{SectEx}
\subsubsection{Two dimensional systems}

\begin{enumerate}

\item{}
\label{case2_1}

\mzc{Let us consider the case of a two - dimensional non - relativistic system. Fermionic fields $\psi$ interact with external non - Abelian $SU(N)$ gauge field $A$, and are in the presence of  the Abelian $U(1)$ gauge field $C$. The field strength of $C$ is of the magnetic type, i.e. we suppose that the only nonzero components of $C_{ij} = \partial_{i}C_j - \partial_j C_i$ are those with $i,j = 1,2$, and that $C_{12} = H = const$.  We then consider $\hat{Q}$ of the form 
\begin{equation}
\hat{Q} =  - \partial_\tau + \mu  - A_0 -  C_0 + \frac{\vec{D}^2}{2 M}, \quad \vec{D} = \vec{\partial} - i \vec{A} - i \vec{C} 	
\end{equation} 
Here $\tau = x^3$ is "imaginary" time coordinate. By $\vec{D}$ we denote the two - dimensional (spatial) covariant derivative. $M$ is mass of the fermions while $\mu $ is chemical potential. Following the calculations presented in \cite{ZW2020} (Appendixes E,F,G) we obtain that the value of $M_3^v$ does not need a regularization (i.e. it remains convergent), and we obtain 
\begin{align}
	\begin{split}
		{M}^v_3&=-\frac{1}{S \,24\pi^2} \int d^2 x\wedge \text{tr}_{D}\left(G^{(0)}\star dQ\star \wedge dG^{(0)}\star\wedge dQ \right) = n 
	\end{split}\label{M3e}
\end{align} 
where $n$ is the number of occupied Landau levels (of the field $H$), i.e. the number of those levels situated below $\mu$. }

\item 

\label{case2_2}
\mzc{Let us modify the case considered above adding Coulomb interactions between fermions with weak coupling constant. Weak interactions within perturbation theory cannot change the value of topological invariant $M_3^v$ as was proven in   
 \cite{ZZ2022}. We come to conclusion that the system exists in the integer QHE states with the same value of $M_3^v$ as without interactions. Notice that the non - perturbative effects of interactions may result in the fractional  QHE, also in the non - Abelian case. However, we do not consider here this possibility. }
 
\item 
 
\label{case2_3}

\mzc{Now let us consider the example, in which the regularization in Eq. (\ref{M3}) is important. Namely, we will be interested in the $2D$ system with relativistic symmetry.
As above, the fermionic fields $\psi$ interact with external non - Abelian $SU(N)$ gauge field $A$, and are in the presence of the non - trivial vierbein field $e_a^\mu$ and the Abelian $U(1)$ gauge field $C$. The field strength of $C$ is again of the magnetic type, i.e.  the only nonzero components of the field strength $C_{ij} = \partial_{i}C_j - \partial_j C_i$ are those with $i,j = 1,2$, and $C_{12} = H = const$.  Operator $\hat{Q}$ has the form 
\begin{equation}
	\hat{Q} =  \frac{1}{2}\sigma^a (|e|e_a^\mu D_\mu + D_\mu |e|e_a^\mu ) - m|e|, \quad \vec{D} = \vec{\partial}- i \vec{A} - i\vec {C}, \quad  {D}_3 = {\partial}_3 -  {A}_0 - {C}_0 + \mu 	
\end{equation} 
Here $\tau = x^3$ is "imaginary" time coordinate. By $\vec{D}$ we denote the two - dimensional (spatial) covariant derivative. $m$ is mass of the fermions while $\mu $ is chemical potential. Consider first the case of constant vierbein. Again, following the calculations presented in \cite{ZW2020} (Appendixes E,F,G) we obtain 
\begin{align}
	\begin{split}
		{M}^v_3&=-\frac{1}{S \,24\pi^2} \int d^2 x\wedge \text{tr}_{D}\left(G^{(0)}\star dQ\star \wedge dG^{(0)}\star\wedge dQ \right) = n 
	\end{split}\label{M3e2}
\end{align} 
Now $n$ is the number of the occupied Landau levels (of the field $H$), i.e. the number of those levels situated below $\mu$. However, the number of those levels is always infinite, and therefore, the above expression is divergent. When we take into account the Pauli - Villars regulators, the value of $M_3^v$ becomes: 
\begin{align}
	\begin{split}
		{M}^v_3&=-\frac{1}{S \,24\pi^2} \sum_{i}C_i\int d^2 x\wedge \text{tr}_{D}\left(G^{(0)}\star dQ\star \wedge dG^{(0)}\star\wedge dQ \right) = \sum_i C_i n_i  
	\end{split}\label{M3e2}
\end{align}  
Here $n_i$ ($i = 1,2,.. $) is the number of occupied Landau levels for the Pauli - Villars regulators with masses $M_i \gg m$, while $n_0 = n$ is the number of occupied Landau levels for the field $\psi$ itself. Suppose that $\mu > 0 $. The number of Landau levels situated below zero does not depend on $M_i,m$ (and thus on $i$) for $ i = 0,1,...$ Because of the condition $\sum_i C_i = 0$ their contributions to $M_3^v$ cancel each other. At the same time the number of occupied Landau levels above zero for the regulators vanishes. We come to conclusion that 
$$
M_3^v = n_{>0}
$$    
where $n_{>0}$ is the number of occupied levels of $\psi$ situated above zero. One can easily find that for $\mu < 0$ the answer will be 
$$
M_3^v = n_{<0},
$$  
 where $n_{<0}$ is the number of vacant Landau levels of $\psi$ situated below zero. }
 
 \item 
 \label{case2_4}
\mzc{ The case considered in the previous item may be complicated adding the slow varying vierbein $e_a^\mu$ and the Coulomb interactions between the fermions with weak coupling constant. As above, using results of \cite{ZZ2022} we come to the conclusion tha the results of Sect. \ref{case2_3} are not modified in this case as well.}
 
\end{enumerate} 

\subsubsection{$3+1$D systems}
\label{SectQHE3D}
\begin{enumerate}
	\item

\mzc{Now let us turn to the $3+1$ D system. We start from the effective low energy field theory of Weyl semimetal. In the proposed model the fermionic fields $\psi$ interact with external non - Abelian $SU(N)$ gauge field $A$. We do not concretize here the origin of this field. Let us just notice that it might appear as an emergent gauge field caused by elastic deformations.  We then consider $\hat{Q}$ of the form 
\begin{equation}
	\hat{Q} = \gamma^\mu i D_\mu +  \gamma^\mu \gamma^5 b_\mu, \quad \vec{D} = \vec{\partial} - i \vec{A}, \quad D_4 = \partial_4 + A_0  - \mu  	
\end{equation} 
Here $x^4$ is "imaginary" time coordinate. By $\vec{D}$ we denote the three - dimensional (spatial) covariant derivative.  $\mu $ is chemical potential, while vector $b_\mu = (\vec{b},0)$ does not depend on coordinates.  Vector $\tilde{M}^{*(v)i}$ of Eq. (\ref{M3}) may be calculated using the formalism developed in \cite{ZW2020}, see also \cite{nissinen2019elasticity}. Namely, we can represent 
\begin{equation}
	\tilde{M}^{*(v)i} = \sum_{j}C_j \int dp_i \tilde{N}^{(j)(i)}_3(P)
\end{equation}
with 
\begin{align}
	\begin{split}
		\tilde{N}^{(j)(i)}_3(P)&=  -\frac{1}{V \,24\pi^2}  \int d^3 x\int_{\Sigma^{(i)}(P)}   \text{tr}_{D}\left(G^{(0)}\star dQ\star \wedge dG^{(0)}\star\wedge dQ \right) 
	\end{split}\label{M3e}
\end{align}
Here integral is over the hyperplane $\Sigma^{(i)}(P)$ in momentum space orthogonal to the $i$ - the axis. The $i$ - the component of the coordinates of all points on this hyperplane are equal to $P$.
Without lost of generality let us consider the case when vector $\vec{{b}}$ is directed along the third axis. The value of  $\tilde{N}^{(0)(3)}_3(P)$ may be calculated easily using the technique developed in \cite{ZW2020}. It is given by 
\begin{equation}
	\tilde{N}^{(0)(3)}_3(P) = \Bigl[ \begin{array}{c}
		1, P \in (-|\vec{b}|, |\vec{b}|)\\
		0, {\rm otherwise}
	\end{array}
\end{equation} 
The other components of  $\tilde{N}^{(j)(i)}_3(P)$ vanish (including those that correspond to the Pauli - Villars regulators).
We arrive at the conclusion that 
 $$
 \tilde{M}^{*(v)i} = 2 b_i
 $$
 and the non - Abelian QHE results in vector current
 \begin{align}
 	\begin{split}
 		\bar{J}^{v(B)}_i&=- \frac{\epsilon_{ijk}}{2\pi^2}b_j E_{k} \label{QHEAGCMEe}
 	\end{split} 
 \end{align}}

\item

\mzc{Now let us consider the case of the system similar to the considered above, but with an extra external $U(1)$ gauge field $C$ of magnetic type (i.e. $C_{ij} = \frac{1}{2}\epsilon_{ijk}H_k$, where $H$ is magnetic field). Operator $\hat{Q}$ now has the form 
\begin{eqnarray}
	\hat{Q} &=& \gamma^\mu i D_\mu   - M_j , \nonumber\\&&\quad \vec{D} = \vec{\partial} - i \vec{A} - i \vec{C},\quad D_4 = \partial_4  -  \mu + A_0 + C_0   	
\end{eqnarray}
Here $M_j$ is the mass parameter (its value $M$ for the Pauli - Villars regulators is assumed to be large, its value $M_0 = m$ for the original fermions may be small), $\kappa$ is constant. 
For simplicity let us consider the case when $\vec{H} \parallel \vec{b}$ while $b^4 = 0$. In this model the Landau levels are formed. Component of momentum along $\vec{b}$ remains good quantum number. We represent again 
\begin{equation}
	\tilde{M}^{*(v)i} = \sum_{j}C_j \int dp_i \tilde{N}^{(j)(i)}_3(P)
\end{equation}
with 
\begin{align}
	\begin{split}
		\tilde{N}^{(j)(i)}_3(P)&=  -\frac{1}{V \,24\pi^2}  \int d^3 x\int_{\Sigma^{(i)}(P)}   \text{tr}_{D}\left(G^{(0)}\star dQ\star \wedge dG^{(0)}\star\wedge dQ \right) 
	\end{split}\label{M3e2}
\end{align} 
As above, we direct the third axis along vector $\vec{b}$. Then for the vierbein not depending on coordinates the  values of $\tilde{N}^{(j)(3)}_3(P)$ may be calculated within an auxiliary $2+1$ D system as the topological invariant 
\begin{align}
	\begin{split}
		\tilde{N}^{(j)(3)}_3(P)&=-\frac{1}{S \,24\pi^2} \int d^2 x\int \text{tr}_{D}\left(\tilde{G}\star d\tilde{Q}\star \wedge d\tilde{G}\star\wedge d\tilde{Q} \right)  
	\end{split}\label{M3e2}
\end{align}
with the $2+1$ D Dirac operator of the form
\begin{eqnarray}
	\hat{\tilde{Q}} &=&  \gamma^\mu i D_\mu -  \gamma^3 P  -  M ,\nonumber\\&& \quad \vec{D} = \vec{\partial} - i \vec{C},\quad D_4 = \partial_4  -  \mu  , 
	\quad \mu = 1,2,4  	
\end{eqnarray} 
Its Weyl symbol obeys $\tilde{Q}_W \star \tilde{G}_W = 1$. Here $P$ plays the role of parameter. Energy levels of the $n$ - th level are given by
\begin{equation}
	E^{(j)}_0 = \pm \sqrt{P^2 + M_j^2 }, \quad
	E^{(j)}_n = {\rm sign}\,(n)\, \sqrt{P^2 + M_j^2 + 2 |n| |B|}, \quad n\in Z, n\ne 0 
\end{equation}
The topological invariant responsible for the QHE is equal to the number of occupied Landau levels (to those with $E^{(j)}(P)\le \mu$):
\begin{equation}
	\tilde{N}^{(j)(3)}_3(P) = \sum_n \theta(\mu- E^{(j)}_n(P))\label{eq737}
\end{equation}
The components of $\tilde{N}^{(j)(i)}_3(P)$ with $i \ne 3$ vanish.
For the purpose of illustration let us consider the case $0< \mu \ll \sqrt{2|H|}$ and $m = 0$. Then the contributions of Landau levels with negative energy for all $j$ are equal (and divergent). These divergent contributions to  $\tilde{M}^{*(v)3}$ cancel each other due to the condition $\sum_j C_j = 0$. We arrive at the contribution of the upper branch of the lowest Landau level of the original fermion for $P \in (-\mu, +\mu)$: 
\begin{equation}
	\tilde{M}^{*(v)3} = 2 \mu\label{eq738}
\end{equation} 
}

\mzo{Notice that the expression of Eq. (\ref{eq737}) is equal to the number of the occupied energy levels situated below the chemical potential. This is integer number, which is robust to the smooth modification of the system. However, the total expression for the $3+1$ D QHE conductivity of the considered system is given by Eq. (\ref{eq738}). It is given by an integral of the topological invariant of Eq. (\ref{eq737}) times $dP$. The result is proportional to the chemical potential, and is not topological invariant. }
 
\end{enumerate}
 
\section{Conclusions and discussion }
In the present paper we proposed a new version of Wigner-Weyl calculus. \mzc{It extends the constructions of \cite{ZW2020,ZZ2022,ZA2023} to the case, when the Wigner transformed two-point fermion Green function belongs to the adjoint representation of the gauge group. In turn, the latter construction utilizes the notion of covariant Weyl symbol of operator proposed long time ago in \cite{vasak1987quantum}.} Basic properties of the so-constructed Wigner-Weyl calculus remain similar to those of the Wigner-Weyl calculus of \cite{ZW2020,ZZ2021,ZA2023}. However, there are important differences that allow us to express responses of vector and axial currents to the non-Abelian field strength through topological invariants. 

Using the developed machinery we calculated responses of axial and vector non-Abelian currents to the external non-Abelian field strength. The former leads us to the non-Abelian version of the CSE while the latter results in the non-Abelian version of the QHE. 
We considered systems with Dirac operator $\hat{Q}$ of rather general form. An important property of the systems considered is their inhomogeneity, which may be caused by spatially varying external fields including those that represent the macroscopic motion of the system \cite{SAZ2024}. \mzc{The obtained results for the non - Abelian CSE and non - Abelian QHE formally repeat the previously known results \cite{ZA2023,Z2024CSE,ZW2020} on the corresponding Abelian effects. In particular, the conductivities of these phenomena are expressed through the topological invariants that are robust to smooth modifications of the systems. Thus in the present paper we extend the calculation of conductivities of the CSE and QHE to the non - Abelian systems using the developed machinery of covariant Wigner - Weyl calculus. We considered several examples of particular models, in which the discussed effects are present. Within these models we calculated the corresponding conductivities, which illustrate our general findings.   }

 The more complicated systems with (in general) non-local and inhomogeneous Dirac operators $\hat{Q}$ may be considered as well. The only restriction on such systems is that the fermion action is to be quadratic in the fermion field. In order to consider \mzc{even} more complicated theories, an interaction with extra fields may be introduced, and the latter fields are to be treated as background ones. 
 
 \mzo{One can see that our expressions for the non - Abelian QHE and CSE look identical to those of the Abelian effects. However,  there is the huge difference behind this. In order to prove that the resulting expressions for the CSE and QHE conductivity are such simple, we had to develop the complicated machinery presented in Sect. 3,4,5. Technically the difference between the Abelian and non – Abelian cases is important. To give an analogy, one can say that this difference is as the difference between the motion of a point particle in special relativity (Minkowski space - time) and general relativity (Riemannian space - time). The action for the particle looks the same (mass times geodesic distance along the worldline). However, behind this there is non – trivial metric in the case of Riemannian geometry, and the presence of Cristofel symbols in the equation of motion. 
 	To proceed, the machinery of Sect. \ref{Sect3}, \ref{Sect4}, \ref{Sect5} is the main result of the present paper. Its complexity and beauty is the main reason for writing the present paper. Notice, however, that a certain piece of an Abelian version of this formalism was developed earlier in \cite{Vasak:1987um}. 
 	In Sect. \ref{Sect6}, and \ref{Sect7} we present only the two particular fields, where this machinery may be applied. As expected, there should be many other fields of mathematical physics, where it can be applied as well. However, already here we see that the results obtained with the aid of the developed formalism are strikingly simple in spite of the complexity of the considered problems. We obtain that the conductivities of QHE in $2+1$ D and CSE in $3+1$ D are given by the topological invariants, while the conductivity of the $3+1$ D QHE is given by an expression that itself might not be a topological invariant, but is related intimately to the topological invariants (similar to the case of the Abelian QHE in Weyl semimetals – see Sect. \ref{SectQHE3D}).
 }

One of the motivations of the present research is to look for the topological meaning of the topological invariant $N_3$ of Eq. (\ref{N3CSE}). It enters expression for the conductivity of chiral separation effect (both Abelian and non - Abelian).  
We investigate the case of background fields which may remain inhomogeneous at infinity, and, therefore, carry non-trivial topology. Namely, the response of axial current to the external magnetic field and to the chemical potential is given by the same topological invariant of  (\ref{N3CSE}). It is expected that this non-Abelian CSE is related to the non-Abelian chiral anomaly in a similar way to the relation between the ordinary CSE and the Abelian chiral anomaly. Establishment of this relation, however, remains out of the scope of the present paper, and will be considered in a forthcoming publication.

Apart from the academic interest, the proposed non-Abelian CSE may appear in real high energy physics. A situation when the non-Abelian magnetic field is present is realized, in particular, inside the non-Abelian vortices. Such vortices exist, in particular, in quark matter in the hypothetical color superconductor phases (see \cite{eto2011dynamics,muneto} and references therein). The closed strings containing color-magnetic flux exist in quark matter in these phases. The non-Abelian CSE then predicts the existence of a non-Abelian axial current along such strings. This may cause certain secondary effects that might play a role in the  physics of quark matter at large baryon chemical potentials.  

The second row of (\ref{QHEAGV}) predicts the appearance of a non-Abelian charge density in the case when both the color-magnetic field and intrinsic magnetic moment co-exist. This may take place inside non-Abelian vortices in the color-superconductor phase when an intrinsic magnetic moment is present in quark matter. The latter may be caused, say, by the ordinary magnetic field coexisting within the non-Abelian vortices with the color-magnetic field. It may also be caused by the macroscopic motion of quark matter. 

\mzc{In addition to investigation of the non-Abelian axial currents, we consider the response of vector non-Abelian current to external (electric) non-Abelian gauge field. In $3+1D$ systems this response is given by an expression, which is given by an integral of a  topological invariant, and it results in the non-Abelian version of the QHE. The emergent non-Abelian gauge fields can occur in condensed matter systems \cite{Volovik2003a} -- including both ultracold atoms and real solids, see also \cite{vyasanakere2011bound,lepori2016double,guo2019two} and references therein. In particular, the emergent non - Abelian gauge fields may appear in Dirac/Weyl semimetals in the presence of elastic deformations. Illustrative example presented in Sect. \ref{SectQHE3D} (item 1) demonstrates that in such systems the non - Abelian QHE is possible. }

\mzc{The two - dimensional non - Abelian QHE may, in principle, be observed in the $2D$ Weyl semimetals  (see \cite{guo2019two} and references therein). }

\begin{appendix}
\begin{section}{Euclidean conventions}
	\label{conventions}	
	In the present paper we use the following conventions concerning vector indices. In relativistic theory there are two types of indices: upper and lower. For the gauge potential these two indices are related by
	$$
	A^\mu = \eta^{\mu \nu} A_\nu ,\quad \mu,\nu = 0,1,2,3
	$$  	
	where $\eta^{\mu\nu} = {\rm diag}(1,-1,-1,-1)$. 
	Consequently, lowering of spatial Minkowski space indices is followed by a change of sign, while lowering of the $0$-th component is not followed by a change of sign.  
	
	In Euclidean space, lowering and lifting of indices does not result in any change of the corresponding components of a vector:
	$$
	A^i = A_i, \quad i = 1,2,3,4
	$$
	
	According to our conventions, Euclidean components of a vector are related to their Minkoswkian components as follows:
	$$
	A^i = A_i = A^\mu = - A_\mu, \quad \mu = i = 1,2,3
	$$
	but
	$$
	A^4 = A_4 = -i A^0 = -i A_0
	$$
	in spite of the following relation between components of the coordinates
	$$
	x^4=x_4 = + i x^0 = + i x_0
	$$
	This is because the Matsubara frequency $\omega = p^4$ is related to the energy ${\cal E} = p^0$ as $p_4 = \omega = - i {\cal E} = -i p_0$. With this identification we obtain for the Dirac operator of a simple system with Lagrangian depending on momentum only:
	$$
	Q(\omega,\vec{p}) = i\omega - {\cal H}(\vec{p})
	$$ 
	with the plus sign in front of $i \omega$. Then 
	$$
	Q(p_4 - A_4,\vec{p} - \vec{A}) = i\omega - A_0 - {\cal H}(\vec{p}-\vec{A})
	$$
	as it should.

	In particular, the above conventions are valid for components of the four-vector of electromagnetic potential.
	
	For the field strength $F$ we have:
	$$
	F_{\mu\nu} = \partial_\mu A_\nu - \partial_\nu A_\mu = \frac{\partial}{\partial x^\mu} A_\nu - \frac{\partial}{\partial x^\nu} A_\mu 
	$$
	We define Euclidean field strength using a similar expression:
	$$
	F^{ij} = F_{ij} = - F_{\mu\nu} = \frac{\partial}{\partial x^i} A_j - \frac{\partial}{\partial x^j} A_i, $$$$ \mu=i = 1,2,3,\quad \nu = j=1,2,3 
	$$ 
	Correspondingly, $F_{ij} = \varepsilon_{ijk} B_k$, where $\vec{B}$ is the magnetic field.
	
	Mixed spatial-time components are:
	\begin{align}
	F^{4j} &= F_{4j} =\frac{\partial}{\partial x^4} A_j -  \frac{\partial}{\partial x^j} A_4 \\
 &=i \partial_{x^0} A_\nu + i \partial_{x^\nu} A_0, \quad \nu = j=1,2,3 
	\end{align}
	and
	$F_{4j} = -i E_j$ since we do not consider a vector potential depending on time.

\end{section}

\section{Magnetic moment and topological invariant responsible for the QHE}

\label{Magnetic}

\mz{We can define the magnetic moment of the system of charged fermions as the response of the thermodynamical potential to external $U(1)$ magnetic field. We assume here that the external electric $U(1)$ field is absent. At the same time the given $U(1)$ is assumed to be a subgroup of the gauge group considered in the main text.  We do not consider in this section the Pauli - Villars regulariation. Extension of the results obtained below to the case when this regularization is added is straightforward.}
	
	\mz{ Let us calculate the response of the partition function to the chemical potential using the developed Wigner-Weyl formalism:
\begin{eqnarray}
		\delta \, {\rm log}\, Z  &=& -i \int d^4 x\, \frac{d^4 p}{(2\pi)^4}\, {\rm tr}_D\, {\rm tr}_G\,{ G}_W\partial_{p_4} Q\,\delta \mu \nonumber\\  &=&-i \int d^4 x\, \frac{d^4 p}{(2\pi)^4}\, {\rm tr}_D\, {\rm tr}_G\, { G}_W(x,p)\bigstar \partial_{p_4} Q(x,p)\,\delta \mu \label{J4}
\end{eqnarray}
In the second line of the above expression we introduced the star product.}

\mz{Next we calculate the response to the external $U(1)$ field strength $\delta {\cal F}_{ij}= \epsilon_{ijk} {\cal B}_k$, where $\cal B$ is the $U(1)$ magnetic field.  This results in
	\begin{eqnarray}
		\delta \, {\rm log}\, Z
		&=& \frac{1}{2}\epsilon_{ijk} \int d^4 x\, \frac{d^4 p}{(2\pi)^4}\, {\rm tr}_D\,{G}_W^{(0)} \star \partial_{p_i} { Q} \star {G}_W^{(0)} \star \partial_{p_j}  {Q}_W \star { G}_W^{(0)}\star \partial_{p_4} { Q}^{(0)}_W\,\delta \mu  \, \delta {\cal B}_{k}\nonumber
	\end{eqnarray}
The integration over imaginary time $x^4$ goes from $0$ to inverse temperature $1/T \to \infty$.	
The response of the thermodynamical potential to variation of the $U(1)$ magnetic field is given by 
$$
\delta \Omega = - \vec{\cal M} \delta \vec{\cal B}
$$
Relation between the thermodynamical potential $\Omega$ and partition function ${\rm log}\, Z = - \Omega/T$ allows us to derive an expression for the derivative of the magnetic moment  $\vec{\cal M}$ with respect to the chemical potential:
	\begin{eqnarray}
		\frac{\partial}{\partial \mu} {\cal M}^i &=&   \frac{1}{2 V} \epsilon_{ijk} \int d^3 r\, \frac{d^4 p}{(2\pi)^4}\, {\rm tr}_D\, { G}_W^{(0)} \star \partial_{p_j} { Q} \star {G}_W^{(0)} \star \partial_{p_k} { Q} \star {G}_W^{(0)} \star \partial_{p_4} { Q}
	\end{eqnarray}
Above, $V$ is the overall volume of the sample.
We can rewrite this expression as follows:
\begin{eqnarray}
	\frac{\partial}{\partial \mu} {\cal M}^i &=&  \frac{1}{4 \pi^2} \, \tilde{M}^{*(v)i}
\end{eqnarray}
where $\tilde{M}^{*(v)i}$, given by Eq. (\ref{M3}),
is a topological quantity in the case that the integrand is free of singularities.}

\end{appendix}


\bibliographystyle{unsrt} 
\bibliography{biblio,cross-ref,wigner3,CSE_MZ,QHE,QFTMacroMotion}

\end{document}